\documentclass{emulateapj}

\newcommand{\be}{\begin{eqnarray}}
\newcommand{\ee}{\end{eqnarray}}
\newcommand{\beq}{\begin{equation}}
\newcommand{\eeq}{\end{equation}}
\def\simless{\mathbin{\lower 3pt\hbox
      {$\rlap{\raise 5pt\hbox{$\char'074$}}\mathchar"7218$}}}
\def\simgreat{\mathbin{\lower 3pt\hbox
      {$\rlap{\raise 5pt\hbox{$\char'076$}}\mathchar"7218$}}} 

\renewcommand{\vec}[1]{\mbox{\boldmath $\displaystyle #1$}}

\newcommand{\grad}{{\mbox{\boldmath $\nabla$}}}

\begin{document}

\title{ Thermal Tides in Short Period Exoplanets }

\author{ Phil Arras\altaffilmark{1} and Aristotle Socrates\altaffilmark{2}   }

\altaffiltext{1}{Department of Astronomy, University of Virginia,
P.O. Box 400325, Charlottesville, VA 22904-4325}

\altaffiltext{2}{Institute for Advanced Study, 
Einstein Drive, Princeton, NJ 08540}
 
\email{ arras@virginia.edu, socrates@ias.edu }

\keywords{(stars:) planetary systems}

\begin{abstract}
Time-dependent insolation in a planetary atmosphere induces a mass
quadrupole upon which the stellar tidal acceleration can exert a
force. This ``thermal tide" force can give rise to secular torques
on the planet and orbit as well as radial forces causing eccentricity
evolution. We apply this idea to the close-in gas giant exoplanets
(``hot Jupiters").
The response of radiative atmospheres is computed in a hydrostatic
model which treats the insolation as a time-dependent heat source,
and solves for thermal radiation using flux-limited diffusion. Fully
nonlinear numerical simulations are compared to solutions of the
linearized equations, as well as analytic approximations, all of which
are in good agreement. We find generically that thermal tide density
perturbations {\it lead} the semi-diurnal forcing. As a result
thermal tides can generate asynchronous spin and eccentricity.
Applying our calculations to the hot Jupiters, we find the following results:
(1) Departure from synchronous spin is significant for hot Jupiters, and 
increases with orbital period.
(2) Ongoing gravitational tidal dissipation in spin equilibrium leads to
steady-state internal heating rates up to $\sim 10^{28}\ {\rm erg\
s^{-1}}$. If deposited sufficiently deep, these heating rates may
explain the anomalously large radii of many hot Jupiters in terms of a ``tidal
main sequence" where cooling balances tidal heating.
At fixed stellar type, planet mass and tidal $Q$,
planetary radius increases strongly toward the star inside orbital periods
$\la 2$ weeks.
(3) There exists a narrow window in orbital period where small eccentricities, $e$,
grow exponentially with a large rate. This window may explain
the $\sim 1/4$ of hot Jupiters which should have been circularized by
the gravitational tide long ago, but are observed to have significant nonzero $e$.
Conversely, outside this window, the thermal and gravitational
tide both act to damp $e$, complicating the ability to constrain the planet's
tidal $Q$.

\end{abstract}

\keywords{planets -- tides}


\section{Introduction}

A number of puzzles have arisen for the gas giant exoplanets
orbiting close to their parent stars, the ``hot Jupiters." 

A large fraction of transiting hot Jupiters are observed to have
radii far larger than the radius of Jupiter, $R_J$,
implying high temperatures deep in the planetary
interior \citep{2000ApJ...534L..97B}. Strong irradiation has been
found to have an insulating effect on planets, slowing their cooling
and contraction \citep{2000ApJ...534L..97B}. This effect can explain
radii $R_p \sim (1.0-1.2) \times R_J$, but is insufficient to explain the radii
of a significant fraction of the population with $R_p \sim (1.2-1.8) \times R_J$ (see
\citealt{2008arXiv0801.4943F} for a recent review). A powerful internal
heat source must be acting to prevent these planets from contracting.

Circularization of the planet's orbit by dissipation of the gravitational
tide has been invoked to explain the small eccentricities of most
hot Jupiters \citep{1996ApJ...470.1187R, 1997ApJ...481..926M}. Using
Jupiter's inferred tidal $Q=10^5-10^6$ \citep{1981Icar...47....1Y}, 
orbits are expected to be circular out to orbital periods $P_{\rm
orb} \la 1\ {\rm week}$. Curiously, a large fraction of the
population that {\it should} have been circularized in $\sim $Gyrs
is not \citep{2004ApJ...610..464D,2008ApJ...686L..29M}. The zero and
nonzero eccentricity planets occupy the same orbital
period range. Suggested explanations are tidal interaction with a
rapidly rotating young star \citep{2004ApJ...610..464D}, perturbations
from other planets, or a large range of tidal $Q=10^5-10^9$  \citep{2008ApJ...686L..29M}. 
In the continued absence of a
detectable perturber with appropriate mass and orbit to account
for the observed eccentricity, a mechanism involving an isolated star
and planet is needed. It is also not clear why planets with presumably similar internal 
structure and orbits should have such different levels of internal 
friction.

In the absence of direct observations, it is commonly assumed that the rotation
of present day hot Jupiters is highly synchronous. This assumption is
motivated by the short ($\sim {\rm Myr}$) synchronization time using
$Q=10^5-10^6$ for the gas giant.  However, despite the short expected
synchronization time, hot Jupiters are particularly susceptible to develop
asynchronous rotation. Gas giants are relatively frictionless compared
to terrestrial planets with $Q \sim 10-100$. Even weak opposing
external torques may compete with the gravitational tide
to produce asynchronous spin.

The proximity of hot Jupiters to their parent star, $\sim 10^2$ times
closer than Solar System giants, implies an insolation stronger by $\sim
10^4$ and stellar tidal forces stronger by $\sim 10^6$. Insolation
and tidal forces may then play a far more important role for close-in
planets. In this paper we discuss ``thermal tide torques", which are
created through the interplay of stellar irradiation and 
gravitational tidal forces.

\citet{1969Icar...11..356G} first applied thermal tide 
torques to explain the observed
rotation rate of Venus. A number of more detailed studies followed
(e.g. Ingersoll and Dobrovolskis 1981). Venus resides in state of 
slow retrograde rotation. Gravitational tides acting
alone would have synchronized Venus' spin to high precision.  
Furthermore, Venus' spin seems to be
in resonance with Earth's orbit. Tides from Earth acting on permanent
quadrupole in Venus should have a negligible effect in comparison to 
solar gravitational torques.
The possibility that torques from solar
gravitational and thermal tides nearly balance that allows a small effect
such as tidal forces from Earth to have an observable effect.

Previous studies of the thermal tide have assumed a thin atmosphere on
top of a solid surface, as appropriate for Venus and Earth. In addition,
they have assumed that the gravitational tide is strongly dissipative ($Q
\sim 10-100$). In applying the thermal tide to close-in gas giant
planets, a difference between our model and those for Venus and Earth
is that we assume an optically thick atmosphere, rather than ground,
completely absorbs the stellar radiation.
We include flux-limited radiative diffusion of heat above and below this
absorbing layer.
We find thermal diffusion 
is crucial in order to accurately estimate the magnitude 
of the thermal tide, particularly for long forcing periods.
In applying our results, we find that the thermal tide 
can affect not only the planet's spin, but also the eccentricity and radius
for close-in planets.

The plan of the paper is as follows. \S\ref{sec:estimates}
contains a qualitative discussion of why thermal tide effects are important.
There, we argue that spin, eccentricity and global 
energetics of hot Jupiters are determined by the competition 
between the thermal and
gravitational tides. \S\ref{sec:TQ} contains a detailed treatment
of atmospheric response to time-dependent insolation, and the resultant
quadrupole moments. Readers wishing to ignore these technical details
should skip this section. Basic equations and geometry are introduced in \S
\ref{sec:eqgeom}, the heating function is expanded in a Fourier series in \S
\ref{sec:fourier}, temperature profiles from the nonlinear simulations
are discussed in \S \ref{sec:tprofile}, quadrupole moments are
computed in \S \ref{sec:Qgrav}, linear perturbation equations are
derived in \S \ref{sec:linear}, and analytic solutions in the high
and low frequency limits are presented in \S \ref{sec:highfreq}
and \ref{sec:lowfreq}, respectively.  Applications to hot Jupiter spin, radius
and eccentricity are in the following sections. Equilibrium spin frequencies are computed
in \S \ref{sec:spineq}. \S \ref{sec:radii} contains tidal
heating rates, and a discussion of the energetics required to power
the observed radii. \S\ref{sec:eccdot} contains eccentricity
growth/decay rates due to both the thermal and gravitational tides. 
Our most detailed results for the simultaneous equilibria of spin rate,
planetary radius and eccentricity are contained in \S \ref{sec:thermal}.
In \S\ref{sec:diffrot} we qualitatively discuss how thermal
and gravitational tidal forces may drive differential rotation
which, in itself, may lead to dissipation.  Summary and
conclusions are presented in \S \ref{sec:summary}.  The Appendices
contains a derivation of the torque, orbital evolution rates, and tidal
heating rates due to the presence of mass quadrupoles of 
arbitrary form, as well as the mass-radius relation and core luminosity.

\section{Basic Idea and Initial Estimates}

Tidal forces become increasingly important with decreasing 
orbital radius. Acting alone, dissipation of the gravitational tide  
leads to synchronous spin and circular orbit for the planet.
Thermal tide torques complicate this simple model,
leading to qualitatively different evolution.
A conceptual introduction and order of magnitude estimates are given below.

\label{sec:estimates}

\begin{figure}
\plotone{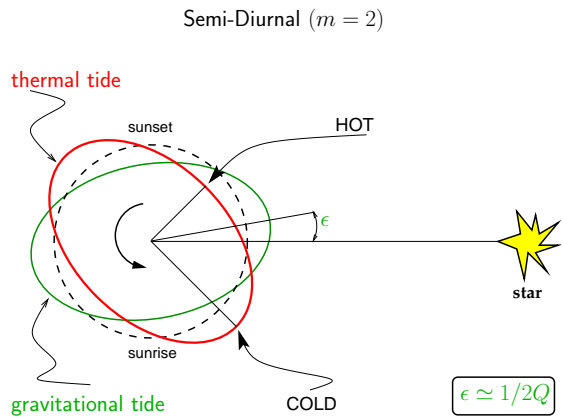}
\caption{ Geometry of the semi-diurnal ($m=2$)
thermal and gravitational tidal density perturbations.
The planet rotates counterclockwise.
In order for the torque to
push the planet away from synchronous spin, the density
perturbation must {\it lead} the line joining the planet and star. }
\label{fig:torque_schematic}
\end{figure}

\subsection{Induced Quadrupoles}

First we summarize the gravitational tide.
The stellar tidal force raises a bulge on the planet, quantified
to lowest order by the quadrupole moment
\be
{\mathcal Q}^{(\rm grav)}\sim \left(\frac{h_t}{R_p}\right)
M_p\,R^2_p\sim M_{\star}\,R^2_p\left(\frac{R_p}{a}\right)^3.
\ee
Here $R_p$ and $M_p$ are the radius and mass of the planet, $M_\star$ is the
mass of the star, $a$ is the semi-major axis of the orbit, and $h_t \sim
R_p (M_\star/M_p)(R_p/a)^3$ is the height of the tide.
In the absence of friction within the planet, the gravitational tide
does not cause secular evolution of the orbital elements or spin. 
The presence of friction causes high tide to lag maximum tidal 
acceleration at noon and midnight. 
For the case of asynchronous spin, this implies that the tidal bulge is 
misaligned with the line joining the planet and star, as pictured
in figure \ref{fig:torque_schematic}.
The stellar tidal force then torques the planet by ``pulling'' 
on the projection of the quadrupole moment ${\mathcal Q^{\rm (grav)}}$
perpendicular to the star-planet line.
This small lag in phase or time is conveniently
parametrized by the tidal $Q$ parameter (e.g. Goldreich and Soter
1965). The effective quadrupole moment contributing to secular 
evolution
is 
\be
{\mathcal Q}^{\rm (grav)}_{\rm sec}\sim \left( \frac{{\mathcal Q}^{\rm (grav)}}
{Q} \right)
\left( \frac{\sigma}{n} \right)
\label{eq:eqQgravest}
\ee 
where $\sigma$ is the tidal forcing frequency and $n$ is the orbital frequency.
The factor $\sigma/n$ ensures the misalignment of the tidal bulge vanishes for synchronous and 
circular orbits (see \S \ref{sec:Qgrav}).
The torque, $N$, on the planet due to this quadrupole moment scales as 
$N \sim n^2 {\mathcal Q}^{\rm (grav)}_{\rm sec}$.

Jupiter's $Q$ is thought to be in the range $Q=10^5-10^6$
based on expansion of Io's orbit to form the Laplace resonance
(e.g. \citealt{1981Icar...47....1Y}).
Several studies have also found that hot Jupiters should have $Q=10^5-10^6$
to explain the circularization of their orbits on Gyr timescales
(e.g. \citealt{2004ApJ...610..477O,2005ApJ...635..688W}). These studies
assume that the gravitational tide is acting alone. Here, we show that
constraints on $Q$ of extrasolar planets are complicated by
the influence of the thermal tide.

Another possibility for tidal phenomena comes from 
intense time-dependent insolation. Time-dependence may arise either through
asynchronous rotation or orbital eccentricity. For circular orbits and
asynchronous spin, the diurnal forcing frequency is
$\sigma = n-\Omega$, where $\Omega$ is the planet's
spin frequency. There is also power
at the higher harmonics $\sigma=m(n-\Omega)$, where $m$ is the azimuthal 
wavenumber.
Since the diurnal component does not generate a quadrupole moment which can 
couple to the 
stellar tidal field, the semi-diurnal
component ($m=2$) dominates the torque on the planet.
The magnitude of the thermal tide
quadrupole is roughly (see \S \ref{sec:highfreq} and \ref{sec:lowfreq})
\be
{\mathcal Q}^{\rm (th)} 
\sim \frac{\Delta M\,R^2_p}{(\sigma\,t_{\rm th})^\delta}
\label{eq:TTQest}
\ee
where $\Delta M$ is the mass of the atmospheric layer heated by time-dependent
insolation,
\be
t_{\rm th} \sim  \frac{\Delta M C_pT}{R^2_p
\,F_{\star}},
\label{eq:tth}
\ee
the exponent $\delta \sim 0.5-1.0$ and $F_\star$ is the flux at the sub-stellar point. 
The thermal time, $t_{\rm th}$, is the time
required for the layer of mass $\Delta M$ to absorb/emit its thermal energy.

One of our principal goals is to show that for the 
atmospheres of interest, the thermal tide bulge leads the sub-stellar
point, opposite the behavior of the dissipative gravitational tide.
\citet{1978Natur.275...37I} made similar arguments for 
Venus based on two assumptions.  First,
the temperature of the absorbing layer lags the time of maximum heating,
similar to our daily experience on Earth. Second, that 
fluid elements remain at roughly constant pressure. Regions of lower temperature then
imply higher density and vice versa, and the high density regions
(``tidal bulges") lead the sub-stellar point.  In our analysis, we
adopt this simplifying ``constant pressure approximation." 
The horizontal readjustment
of mass necessarily requires zonal winds. 
In this paper we do not solve for
the wind structure; see \citet{1980Icar...41....1D} for
the wind structure resulting from a similar calculation applied to
Venus. 

Both the gravitational and thermal tidal bulges
can be torqued by the stellar tidal field. The angular momentum 
evolves due to the net torque.
In spin equilibrium, the thermal and gravitational tide torques
balance, in general leading to an asynchronous spin state. We estimate
the ratio of quadrupole moments to be (eq.\ref{eq:eqQgravest} and \ref{eq:TTQest})
\be
\frac{{\mathcal Q}^{\rm (th)}}{{\mathcal Q}^{\rm (grav)}_{\rm sec}}
& \sim  & \left( \frac{n}{\sigma} \right)^{1+\delta}
\left( \frac{\Delta M}{M_\star} \frac{a^3}{R_p^3}\right)
\left( \frac{Q}{(n t_{\rm th})^\delta } \right)
\nonumber \\ & \simeq & 
\left( \frac{n}{\sigma} \right)^{1+\delta}
\left( \frac{\Delta M}{10^{-8} M_p} \right)
\left( \frac{M_p}{10^{-3}M_\star} \right)\
\nonumber \\ & \times & \left( \frac{a}{100R_p} \right)^3
\left( \frac{Q}{10^5} \right)
\frac{1}{\left( n t_{\rm th} \right)^\delta } .
\label{eq:quad_compar}
\ee
This order of magnitude estimate shows that for parameters characteristic
of the hot Jupiters the thermal and gravitational tide effects can
be comparable. Setting eq.\ref{eq:quad_compar} equal to unity allows
one to estimate the equilibrium spin frequency since $\sigma \propto
n-\Omega$.

\subsection{ Orbit, spin, and thermal equilibrium }

Gravitational tidal dissipation converts rotational and orbital energy into heat.
When the gravitational tide alone acts, the end state is synchronous
spin and a circular orbit. As torque equilibrium is realized
in $\sim {\rm Myr}$, over Gyr timescales the only source
of tidal heating is finite eccentricity.  In the absence
of an external perturbation continually pumping the eccentricity,
tidal heating is limited by the initial energy reservoir in the orbit
$\sim GM_\star M_p e^2/2a$, which may be comparable to planet's binding
energy for initially eccentric orbits \citep{2001ApJ...548..466B}.

Planets with anomalously large radii but small eccentricity, such
as HD 209458b with $e \simeq 0.01$ \citep{2005ApJ...629L.121L}, are
difficult to explain with the gravitational tide alone since the current
eccentricity implies negligible tidal heating. In order to explain the
large radius with transient tidal heating one has to invoke that
heating ended recently, and we are observing the planet before it has had time
to contract. Since contraction times for radii $(1.2-1.8) \times R_J$
are much shorter than a Gyr, this argument requires that we
observe these systems during a special time in their evolution.

Inclusion of the thermal tide qualitatively changes this picture.
A steady state may be reached in which thermal and gravitational
tide effects balance one another. In this scenario, gravitational
tidal dissipation continues to act in the equilibrium state, due
to asynchronous spin and/or eccentricity. The thermal tide forces
are not inherently dissipative. However differential rotation set up in
the atmosphere by opposing torques at different depths may cause
significant heating.

Tangential forces applied to the thermal tide quadrupole torque the
planet away from synchronous spin. In equilibrium, the thermal
and gravitational tide torques on the planet are in balance, setting
the equilibrium spin rate. The timescale to attain equilibrium spin
is shorter than the Gyrs age for orbital periods up to several
months. Radial forces from the thermal tide quadrupole act to alter
the orbital eccentricity. If both the thermal and gravitational tides
act to circularize the orbit, the equilibrium state is again zero
eccentricity. However, the thermal tide can also pump eccentricity,
opposing the gravitational tide. In that case, a balance of thermal
and gravitational tides implies nonzero values for the equilibrium
eccentricity. The timescale to attain equilibrium eccentricity is shorter
than the age for planets with orbital periods shorter than $\sim $
1 week. Lastly, in the absence of tidal heating, planets cool and contract.
If tides deposit heat deep in the convective interior of the planet, 
a thermal equilibrium is possible in
which heating of the core is balanced by outward heat loss at the 
radiative-convective boundary. We find thermal equilibrium is achieved
on a timescale shorter than the age for orbital periods $\la 1$ week (see
section \ref{sec:thermal}).

The thermal tide torque is exerted on a layer near the photosphere
of the stellar radiation. The depth dependence of the gravitational
tide torque is uncertain as the nature of gravitational tidal
dissipation is complicated and not yet fully understood. For
instance, turbulent viscosity damping would mainly occur just
below the radiative-convective boundary \citep{1977Icar...30..301G,
2005ApJ...635..688W}, shear layers due to ``wave attractors" reside
deep in the convective core \citep{2004ApJ...610..477O, 2008arXiv0812.1028G}, and upward
propagating waves generated at the tropopause may break above the
photosphere \citep{2004ApJ...610..477O}. If the thermal and gravitational
torques are not exerted at the same depth, differential rotation is induced, distinct from the usual thermally-driven circulation
patterns. For simplicity, we defer effects due to differential rotation
until \S \ref{sec:diffrot}.

\subsection{ Energetics, and implications for the lightcurve and spectrum  }

The stellar tidal acceleration creates regions of high and low gravitational potential
in the longitudinal direction. Insolation heats the fluid, and this
heat can be tapped to perform work against the stellar tidal field by 
causing pressure gradients which move fluid from regions of low to high potential.
The ultimate source of energy for tidal heating driven by the thermal
tide is then the stellar radiation field.
The rate of work done by the thermal tide is
\be
\dot{E}^{\rm (TT)}_{\rm work} & \simeq & - \Delta M \frac{\sigma}{m} \oint 
\left(- \grad U \right)\cdot d{\bf l}
\simeq - \frac{\sigma}{m} N^{\rm (TT)},
\ee
where $U$ is the tidal potential, $\sigma=2(n-\Omega)$, $m=2$, and
 the integral is taken along a fluid trajectory $d{\bf l}=R_p\,d\phi\,
\hat{\bf \phi}$. Here
$N^{\rm (TT)} \simeq \Delta M\oint \left(- \grad U \right)\cdot d{\bf l}$ 
is the thermal tide torque on the heated layer.
In torque equilibrium, the thermal and gravitational tide torques balance
giving $N^{\rm (GT)}=-N^{\rm (TT)}$. The heating rate due to the
gravitational tide is then
\be
\dot{E}^{\rm (GT)}_{\rm heat} & = & \frac{\sigma}{m} N^{\rm (GT)}
= - \frac{\sigma}{m} N^{\rm (TT)} = \dot{E}^{\rm (TT)}_{\rm work},
\ee
showing explicitly that the work done on the atmosphere by the thermal tide is converted into
heat by the gravitational tide in steady state.

Since the work done on the atmosphere is ultimately powered by the stellar radiation,
the maximum power input is given by the rate of absorption of stellar flux, 
\be
\dot{E}_{\rm max} & = & L_\star \left( \frac{R_p}{2a} \right)^2 \left( 1-e^2 \right)^{-1/2}
\nonumber \\ & = & 
9 \times 10^{28}\ {\rm erg\ s^{-1}} \left( \frac{R_p}{R_J} \right)^2
\left( \frac{L_\star}{L_\odot} \right) 
\nonumber \\
& \times & 
\left( \frac{M_\odot}{M_\star} \right)^{2/3} \left( \frac{4\ {\rm days}}{P_{\rm orb}} \right)^{4/3}
 \left( 1-e^2 \right)^{-1/2}
\ee
where $L_\star$ is the stellar luminosity and $P_{\rm orb}=2\pi/n$ is the orbital period. We will show
in \S \ref{sec:radii} that the efficiency of converting stellar flux to heat by gravitational
tidal dissipation is as high as $\sim 1-10\%$, and increases toward the star.

Some of the heat deposited by insolation is converted into work, leading to less thermal energy immediately re-radiated back out into space.  In the (unphysical) limit of 100\% efficient conversion of heat to work, and no subsequent conversion of kinetic energy back into heat, the atmosphere would be far cooler than the equilibrium temperature, and the thermal emission from the planet would be solely due to the flux coming out from the core. In torque equilibrium, the work done on the atmosphere by the thermal tide is converted into heat by the gravitational tide. We presume the gravitational tide dissipation occurs in the convective core, and entropy is efficiently mixed throughout. The temperature profile as a function of depth, latitude and longitude may then differ markedly from the case with no tidal heating.

First consider the change in the lightcurve for thermal radiation. For small flux from the core, and short thermal time in the absorbing layer, there will be a large day-night temperature difference. If, for example, 1\% of the insolation is deposited in the core by the gravitational tide, then the day-night temperature ratio is $(F_{\rm core}/F_\star)^{1/4} \sim 0.01^{1/4} \sim 0.3$, where $F_{\rm core}$ is the flux emerging from the convective core. For small $F_{\rm core}$, the only possibility of high temperature on the night side is redistribution of heat by zonal winds. Tidal dissipation generating a large
$F_{\rm core}$ provides an alternative means to reduce the day-night temperature contrast (see figure \ref{fig:lum_vs_rad} for the luminosity exiting the core versus planetary radius).

A change in the vertical distribution of heat sources also affects the vertical temperature profile at the depths where the spectrum is formed. Depositing heat in the core rather that at the photosphere leads to a radiative flux more constant with depth near the photosphere. This alteration of the temperature profile may in principle be imprinted on the spectrum.

\section{ Temperature profiles and quadrupole moments }
\label{sec:TQ}

In this section we describe the model for time-dependent
thermal forcing of
hot Jupiter atmospheres due to asynchronous rotation and/or an 
eccentric orbit, as well as the
the resultant gravitational forces on the perturbed atmosphere. The
full nonlinear equations are presented, and then approximated as a
time-independent background and harmonic perturbations. Analytic solutions
are developed for the background and perturbations, and compared to the
full nonlinear solutions.

\subsection{ Equations and geometry }
\label{sec:eqgeom}

We consider a planet and star of mass and radius $(M_p,R_p)$
and $(M_\star,R_\star)$, respectively.  The planet orbits with separation
$D(t)$ and true anomaly $\Phi(t)$ which we will treat as a nearly
Keplerian orbit with semi-major axis $a$, eccentricity $e$ and mean
motion $n=[G(M_\star+M_p)/a^3]^{1/2}$. We consider an atmosphere in
uniform rotation with rate $\Omega$, and the spin and orbital angular
momentum aligned.
Differential rotation will be discussed in \S
\ref{sec:diffrot}. We will work in a non-rotating coordinate system
whose origin is at the center of the planet, and whose axes are fixed
with respect to distant observers.  Spherical polar coordinates
$(\theta,\phi)$ are used, where the star orbits at the equator with a 
colatitude $\theta=\pi/2$ and longitude $\phi=\Phi(t)$, and a fixed
point on the planet rotates with angular frequency $\dot{\phi}=\Omega$.
The derivative comoving with the planet is then $d/dt=\partial/\partial
t + \Omega \partial/\partial \phi$. The cosine of the angle $\chi$
between the vertical and the vector to the star is $\cos\chi=\sin\theta
\cos(\phi-\Phi)$. The day side is over the angular range $-\pi/2 \leq
\phi-\Phi \leq \pi/2$.  Let $z$ be the altitude above some appropriate
reference level and $y(z)=\int_z^\infty dz' \rho(z')$ the mass column
above that altitude. The atmosphere is treated as being both thin and
in hydrostatic balance, hence the pressure $P=gy$, where the surface
gravity is $g=GM_p/R_p^2$.

We use an approximate treatment of radiative transport in which
the time-dependent insolation is treated as a specified heat source.  We approximate
the transfer of thermal radiation by flux-limited diffusion. 
For the thermal radiation,
we employ the solar composition ``condensed" phase Rosseland opacities
of \citet{2001ApJ...556..357A}, which include the effect of grains
in the equation of state, but ignores their opacity, as is appropriate
if the grains have rained out to higher depth. 
There is a prominent dip in $\kappa$ along isobars centered around $T=2000\ {\rm K}$. This feature will be apparent in our
numerical results.

We use the equation of state from \citet{1995ApJS...99..713S}, with 70\% hydrogen and 
30\% helium by mass. At low density, this is an ideal gas equation of state 
$P=\rho k_b T/\mu m_p$, where
$\rho$ is the mass density, $T$ is the temperature, $m_p$ is the proton
mass, $k_b$ is Boltzmann's constant, and $\mu \simeq 2.4$
is the mean molecular weight for a mixture of molecular hydrogen and helium.

The temperature and flux are found by solution of the heat equation
\be
C_p \frac{d T}{d t} & = & \frac{\partial F}{\partial y}
+ \epsilon
\label{eq:fullheat}
\ee
and the equation for flux-limited diffusion
\be
F & = & \frac{16\sigma_{\rm sb} T^3\Lambda}{\kappa} \frac{\partial
T}{\partial  y}.
\label{eq:fullflux}
\ee
Here $C_p\simeq 7k_b/2\mu m_p$ is the specific heat per gram,
and $\epsilon$ is the heating rate per gram. Since column,
or equivalently pressure, is assumed constant for fluid elements, the
thermodynamic relation $ds/C_p=dT/T - \nabla_{\rm ad}dP/P$ is simplified
to $Tds=C_pdT$. We ignore vertical fluid motion relative to constant
pressure surfaces, as well as horizontal motions relative to the mean
rate $\Omega$.  The quantity $\Lambda$ is the flux limiter, which allows
a smooth transition between the optically thick regime, $\Lambda=1/3$,
and the optically thin regime $\Lambda \rightarrow 0$. For
convenience, we use the limiter prescription of \citet{1981ApJ...248..321L}
\be
\Lambda & =& \frac{2+R}{6+3R+R^2}
\ee
where $R=(4/\kappa)|\partial \ln T/\partial y|$ is the ratio of
photon mean free path to temperature scale height.  The boundary
conditions are (1) the radiation free streams at small optical
depths, $F\simeq 4\sigma_{\rm sb} T^4$, and (2) $F \rightarrow F_{\rm
core}$ at large $y$, where $F_{\rm core}$ is the thermal flux that emerges
from the convective core. 

We refer to direct solutions of eq.\ref{eq:fullheat} and 
\ref{eq:fullflux} as ``nonlinear solutions", as temperature changes
are not assumed to be small. The nonlinear solutions will be compared
to solutions of the linearized equations which are valid only for 
small temperature changes.

To numerically solve eq.\ref{eq:fullheat} and
\ref{eq:fullflux}, we adapted the code of
\citet{2005ApJ...628..401P}.  We use second-order finite difference in
$y$ and backward difference in $t$ for stability.  Typically 128 $y$
grid points were used, but selected results were checked at higher
resolution and found to be accurate. The time step is set so that the
average temperature at each grid point changes by $\la 10^{-3}$ over the
time step.  A background model, which we discuss below, 
is used as an initial condition, and
the initial time taken to be noon ($\phi=\Phi$). 
Transients due to the
initial conditions are observed to relax over a few forcing periods.

\subsection{Time-dependent insolation and Fourier components}
\label{sec:fourier}

The time-dependent heating rate per unit mass is given by Beer's Law \citep{Liou2002}
\be
\epsilon(y,\theta,\phi,t) & = & 
\kappa_\star F_\star(D) \exp\left( - \frac{\kappa_\star y}{\cos\chi} \right) \Theta(\cos\chi)
\label{eq:epsilon}
\ee
where the constant $\kappa_\star$ 
specifies the column ($y \sim \kappa_\star^{-1})$ at which the
stellar radiation is absorbed, and $F_\star(D)=\sigma_{\rm sb} T_\star^4 (R_\star/D)^2$ 
is the bolometric stellar
flux at the subsolar point on the planet. We present
numerical results only for a solar-like star with $M_\star=M_\odot$, 
$T_\star=5780\ {\rm K}$ and $R_\star=R_\odot$.  However, 
our analytic analysis allows for rescaling to stars of different mass and 
luminosity.
The step function $\Theta(\cos\chi)$
enforces heating only on the day side. In radiative equilibrium, 
the thermal flux exiting the top of the atmosphere is
\be
F(y=0,\theta,\phi,t) & = & F_{\rm core} + \int_0^\infty dy\
\epsilon(y,\theta,\phi,t)
\nonumber \\ & = & 
F_{\rm core} +  \cos\chi F_\star \Theta(\cos\chi).
\label{eq:fullF0}
\ee
We take the upper limit of integration to be $y=\infty$ since 
$y=\kappa_\star^{-1}$ is a shallow layer in the atmosphere, and the
integral converges exponentially.

The gravitational tide can only couple to fluid perturbations with
quadrupolar angular dependence and a specific harmonic time
dependence. Therefore, we now expand the insolation 
in terms of a Fourier series in longitude and time.
Since $\cos\chi$ depends on only the combination 
of longitudes $\psi=\phi-\Phi$, we write 
\be
\epsilon(y,\theta,\phi,t) & = & \kappa_\star F_\star(D) \sum_{m=-\infty}^\infty 
g_m \left( \frac{\kappa_\star y}{\sin\theta} \right) e^{im(\phi-\Phi)}
\ee
where the integral
\be
g_m(s) & \equiv & \frac{1}{2\pi} \int_{-\pi/2}^{\pi/2} d\psi\ e^{-im\psi-s
\sec\psi}
\label{eq:gm}
\ee
contains a contribution only from the day side.
We compute the integrals $g_m(s)$
numerically. The $m=0$ term is the average over longitude.
For eccentric orbits, using 
$F_\star(D)=F_\star(a)(a/D)^2$, the time-dependence can be expanded as a sum of harmonic terms using
the Hansen coefficients defined in eq.\ref{eq:Hansen} of appendix \ref{sec:appendix1}
\be
\left( \frac{a}{D} \right)^2 e^{-im\Phi} = \sum_{k=-\infty}^\infty X^{1m}_{k}(e) e^{-iknt}.
\ee
The heating rate in eq.\ref{eq:epsilon} can then be written in a Fourier series in both 
longitude and time as
\be
\epsilon(y,\theta,\phi,t) & = & \kappa_\star F_\star(a) \sum_{mk} X^{1m}_k(e)
g_m \left( \frac{\kappa_\star y}{\sin\theta} \right) e^{im\phi-iknt}.
\label{eq:epsmk}
\ee
The $(m,k)$'th term in the series has a
forcing frequency $\sigma_{mk}=kn-m\Omega$ in the co-rotating frame and
$kn$ in the inertial frame. 
The $(m,k)=(0,0)$ term is the time and longitude averaged heating rate, 
which we denote as 
\be
\epsilon(y,\theta)=F_\star(a)\kappa_\star X^{10}_0(e) g_0(\kappa_\star y/\sin\theta).
\label{eq:epsbg}
\ee
Since $X^{1m}_0(e)=\delta_{m0}(1-e^2)^{-1/2}$, there are no time-independent ($k=0$)
non-axisymmetric ($m\neq 0$) heating terms. Axisymmetric ($m=0$) time-dependent
heating does occur for eccentric orbits, due to the radial force.
For the time-independent background, $\int_0^\infty dx\ g_0(x)=1/\pi$ leads to a background surface flux 
\be
F(y=0,\theta) & = & F_{\rm core} + \int_0^\infty dy\ \epsilon(y,\theta)
\nonumber \\ & = & 
 F_{\rm core} + \frac{1}{\pi} F_\star(a) X^{10}_0(e) \sin\theta
\ee
and thermal  luminosity 
\be
L & = & R_p^2 \int d\Omega F(y=0,\theta)
\nonumber \\
& = &  4\pi R_p^2 \left( F_{\rm core} + \frac{X^{10}_0(e)}{4} F_\star(a) \right)
\ee
in thermal radiation from the planet.
We will use eq.\ref{eq:epsbg} to define background models, on top of which we solve for 
perturbations. We define the time-dependent perturbation to the heating rate as
\be
\delta \epsilon(y,\theta,\phi,t) & = & 
\kappa_\star F_\star(a) \sum_{m,k\neq 0} X^{1m}_k(e)
g_m \left( \frac{\kappa_\star y}{\sin\theta} \right) e^{im\phi-iknt}.
\ee
For $e=0$,
$X^{\ell m}_k(0)=\delta_{mk}$ enforces $k=m$, and the forcing frequency
becomes $\sigma_{mm}=m(n-\Omega)$, where $2\pi/|n-\Omega|$ is the diurnal
forcing period.

The values of $g_m(s)$ in the heating function decrease with increasing $m$.
While the diurnal ($m=\pm 1$) terms dominate the observable flux variation,
they do not contribute to the torque
since the tidal potential has no dipole component. The quadrupole moments responsible
for spin and orbital evolution
are dominated by the semi-diurnal ($m=0,\pm 2$) components.  

\subsection{ Discussion of temperature profiles }
\label{sec:tprofile}

\begin{figure}
\plotone{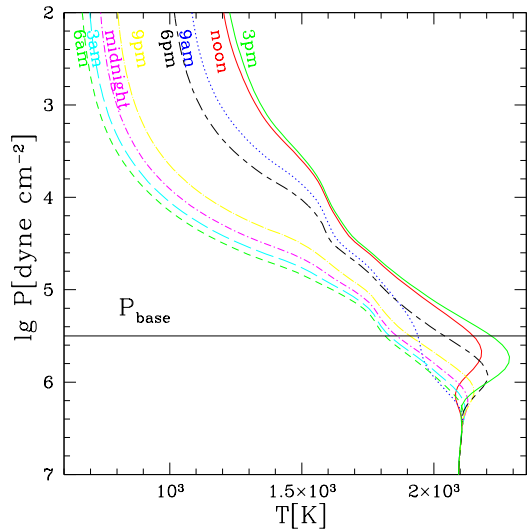}
\caption{ Temperature versus pressure over one diurnal forcing period for ``deep" forcing at the
equator ($\theta=\pi/2$).
Note the lag between maximum forcing (at noon) and maximum temperature (between noon and 3pm),
and also the pronounced ``bump" near the base of the heated layer 
at $P_{\rm base}=g/\kappa_\star=10^{5.5}\ {\rm
dyne\ cm^{-2}}$. This plot was produced using the nonlinear simulations for a circular orbit
with $P_{\rm orb}
=4\ {\rm days}$, super-synchronous rotation $P_{\rm spin}=3\ {\rm days}$,
$\kappa_\star=10^{-2.5}\ {\rm cm^2\ g^{-1}}$,
$g=10^3\ {\rm cm\ s^{-2}}$, $F_{\rm core}=10^4\ {\rm erg\ cm^{-2}\ s^{-1}}$ and a solar type
star. Noon is the time of maximum heating. }
\label{fig:TP_fullday_deep_Porb=4_Pspin=3}
\end{figure}

\begin{figure}
\plotone{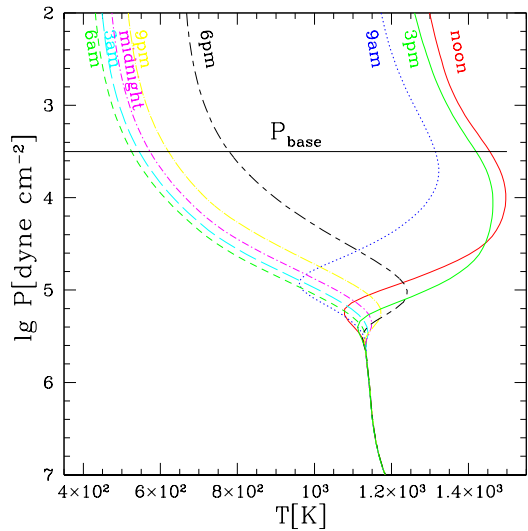}
\caption{ Same as figure \ref{fig:TP_fullday_deep_Porb=4_Pspin=3} but with ``shallow" forcing
$\kappa_\star=10^{-0.5}\ {\rm cm^2\ g^{-1}}$.
The base of the heated layer is now at $P_{\rm base}=g/\kappa_\star=10^{3.5}\ {\rm
dyne\ cm^{-2}}$. The temperature perturbations diffuse significantly below the base of the layer
directly heated.
Here the forcing period, $2\pi/\sigma$, is large in comparison to the thermal
time $t_{\rm th}$ at $y \la \kappa_\star^{-1}$ giving
temperature maximum in phase with the maximum heating at noon.  A
thermal diffusion wave penetrates to large depths, and lags in phase with
respect to the temperature at  $y \la \kappa_\star^{-1}$.  This lag in phase
leads to a torque that attempts to increase the spin of the planet.
}
\label{fig:TP_fullday_shallow_Porb=4_Pspin=3}
\end{figure}

Figures \ref{fig:TP_fullday_deep_Porb=4_Pspin=3} and
\ref{fig:TP_fullday_shallow_Porb=4_Pspin=3} show temperature versus
pressure over one diurnal forcing period for a planet in circular
orbit around a solar-type star. The orbital period is $P_{\rm orb}=4\
{\rm days}$ and the spin period $P_{\rm spin}=3\ {\rm days}$, giving
a diurnal forcing period of $12$ days.  Each curve is labeled by
phase in the heating cycle, where noon means heating is maximum,
etc.  Figure \ref{fig:TP_fullday_deep_Porb=4_Pspin=3} illustrates
a heating function which extends ``deep", down to a base pressure
$P_{\rm base}=g/\kappa_\star=10^{5.5}\ {\rm dyne\ cm^{-2}}$, while figure
\ref{fig:TP_fullday_shallow_Porb=4_Pspin=3} shows ``shallow" heating
down to $P_{\rm base}=g/\kappa_\star=10^{3.5}\ {\rm dyne\ cm^{-2}}$. Naively
one may have expected the temperature to change significantly only
above $P_{\rm base}$. This is approximately true for the deep heating
case. The shallow heating case, however, shows large temperature
perturbations extending a factor of $\sim 30$ deeper in pressure than
$P_{\rm base}$. Heating below $P_{\rm base}$ is due to 
diffusion, rather than direct absorption of stellar radiation. 
We will show that there are two limits for the forcing
period. In the ``high frequency limit", appropriate for planets at
large separation from the star, the temperature perturbations are large
only for depths $P \la P_{\rm base}$.  In the ``low frequency limit",
appropriate for close-in planets, temperature perturbations have time
to diffuse down below $P_{\rm base}$, so that the star torques a layer
deeper than that directly heated by the stellar radiation. 

The ``bump" in the temperature profiles in figure
\ref{fig:TP_fullday_deep_Porb=4_Pspin=3} at $P=10^{5.0}-10^{6.2}\ {\rm
dyne\ cm^{-2}}$ is evidence of an inwardly propagating thermal diffusion
wave.  This bump is also seen in temperature profiles in figure 5 of
\citet{2008arXiv0809.2089S}, although the authors do not identify this
feature as such. \citet{2008arXiv0809.2089S} have a far more detailed
solution method, including nonlinear three-dimensional hydrodynamics,
angle and frequency dependent radiative transfer, and detailed chemical
abundances and monochromatic opacities.  That our solutions are in
qualitative and rough quantitative agreement in the region of interest
lends confidence to our results.  

Another important comparison to the more rigorous results of
\citet{2008arXiv0809.2089S} concerns the lag in the temperature profile,
shown in their figure 3. Although their simulation is of a synchronous
planet, they generate a super-rotating equatorial jet which mimics the
super-synchronous rotation in our results. Those authors indeed find
the temperature reaches a maximum {\it eastward} of the subsolar point;
maximum temperature lags maximum heating. This lag is especially prominent
at large depth (their bottom panel). We regard this as strong evidence that,
while inclusion of fluid motion may alter our results quantitatively,
the qualitative result that temperature lags, due to thermal inertia,
is in agreement with \citet{2008arXiv0809.2089S}. A crucial omission
in their simulations is that the tidal force from the star is not
included. We argue, based on the temperature lag seen in their figure
3, that the torque on this thermally-generated quadrupole will act to
generate asynchronous spin.

\begin{figure}
\plotone{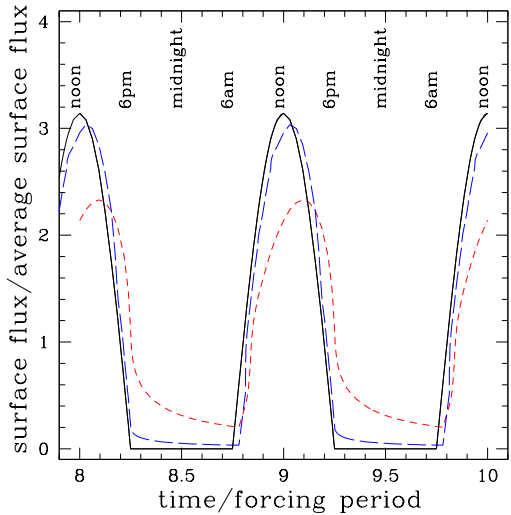}
\caption{ Thermal flux at the surface versus time for the cases
shown in figures \ref{fig:TP_fullday_deep_Porb=4_Pspin=3}
and \ref{fig:TP_fullday_shallow_Porb=4_Pspin=3}. These simulations are at the equator
($\theta=\pi/2$), flux is normalized by the average surface flux
$F_{\rm core} + F_\star(a)/4$, and time is in units of diurnal forcing period (12 days).
The solid black line is the
outgoing flux if no net heat is absorbed or emitted by the atmosphere
(eq.\ref{eq:fullF0}). The blue long-dashed line is the shallow heating case
(figure \ref{fig:TP_fullday_shallow_Porb=4_Pspin=3}). The red short-dashed line
is the deep heating case (figure \ref{fig:TP_fullday_deep_Porb=4_Pspin=3}).
}
\label{fig:Fvst}
\end{figure}

Figure \ref{fig:Fvst} shows the thermal flux exiting the
surface, in units of the average thermal flux, for the two cases
shown in figures \ref{fig:TP_fullday_deep_Porb=4_Pspin=3} and
\ref{fig:TP_fullday_shallow_Porb=4_Pspin=3}. The shallow heating
case shows larger flux variations and smaller lag time in comparison
to the deep heating case. In both cases, the maximum in thermal
emission {\it lags} the maximum in heating (at noon) due to thermal
inertia. The Spitzer lightcurves for HD 189733b show maximum thermal
emission just before secondary eclipse.  Using the geometry from figure
\ref{fig:torque_schematic}, maximum emission occurs just before secondary
eclipse for a planet with {\it super-synchronous} rotation, in order
that the temperature reaches maximum to the {\it east} of the subsolar
point, after it has passed through noon.  A similar effect would occur
for a super-synchronous zonal wind, as in
the model of \citet{2008arXiv0809.2089S}.

One technical point about flux-limited diffusion is that it produces
temperature profiles which increase inward at small optical depth. 
The ``greenhouse" case shown in figure
\ref{fig:TP_fullday_deep_Porb=4_Pspin=3} is captured in our solutions.
Recall that greenhouse heating is due to $\kappa_\star \ll \kappa$; stellar
irradiation deposits heat at large optical depth, leading to
a temperature increase by a factor $\sim (\kappa/\kappa_\star)^{1/4}$ above
the skin temperature $(F_\star/4\sigma_{\rm sb})^{1/4}$. There has been recent interest
in shallow heating, $\kappa_\star \gg \kappa$, due to TiO/VO absorption at
$P \sim $ mbar, producing an inwardly decreasing temperature profile,
the ``stratosphere" case \citep{2003ApJ...594.1011H,2008ApJ...678.1419F}. 
To model this effect would require angle
and frequency-dependent solution of the transfer equations, beyond
the scope of this paper. We note, however, that below $P_{\rm
base}$, the stratosphere case is expected to have temperature $\sim
(F_\star/\sigma_{\rm sb})^{1/4}$, similar to our results with flux-limited diffusion.
Since the torque is applied well below $P_{\rm base}$, we expect our
results to be qualitatively, and perhaps quantitatively correct.

\subsection{ Quadrupole moments }
\label{sec:Qgrav}

Appendix \ref{sec:appendix1} contains a derivation of the secular changes in $e$, $a$ and $\Omega$ due to
a quadrupole moment. In addition, we derive the change in total spin plus orbital energy
due to this quadrupole moment.  Here, we outline the method of computation for 
the thermal tide
quadrupole moment, as well as review Darwin's theory of secular evolution for gravitational
tides.

\begin{figure}
\plotone{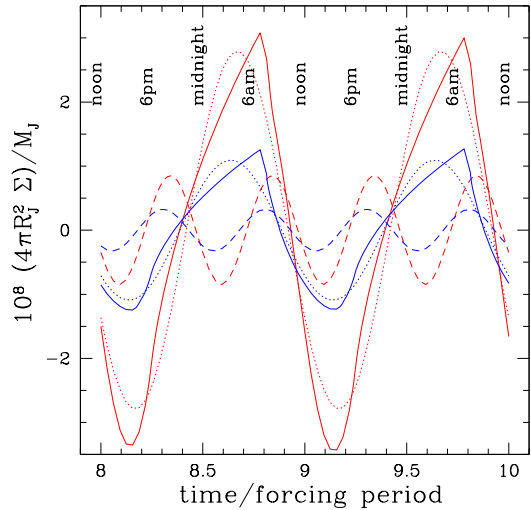}
\caption{ Effective column $\Sigma$ (eq.\ref{eq:Sigma}), converted into units of fractional mass of the
planet, for the deep heating
case (red, figure \ref{fig:TP_fullday_deep_Porb=4_Pspin=3})
and the shallow heating case (blue, figure \ref{fig:TP_fullday_shallow_Porb=4_Pspin=3}).
The solid lines are the full numerical result. Fourier transform of the solid lines gives the diurnal component
(dotted lines) and the semi-diurnal component (dashed lines).
}
\label{fig:colvst}
\end{figure}

The ``thermal tide" force results from the mass multipole moments induced in the 
atmosphere by time-dependent insolation. 
The multipole moments are defined in terms of the density field
that is the solution to eqs. \ref{eq:fullheat} and \ref{eq:fullflux}.  
We have
\be
{\mathcal Q}^{\rm (TT)}_{\ell m}(t) & =& \int d^3x\ \rho(\vec{x},t) 
r^\ell \ Y_{\ell m}^*(\theta,\phi).
\label{eq:Qlm1}
\ee
The density $\rho(\vec{x},t)$
is computed from the temperature profile $T(\vec{x},t)$ using the
ideal gas law and hydrostatic balance $P=gy$.
Since $r^\ell \rho(\vec{x},t)$ is a real quantity, and $Y_{\ell m}^*=(-1)^m Y_{\ell, -m}$, the
moments must satisfy
$\left( {\mathcal Q}^{\rm (TT)}_{\ell m}\right)^* = {\mathcal Q}^{\rm (TT)}_{\ell, -m} (-1)^m$.
As the static and azimuthally symmetric background model has zero quadrupole moment, it can
be subtracted off the integrand. We then make a change of integration variable from
radius $r$ to column $dy=-\rho dr$, and approximate $r \simeq R_p$,
yielding the expression 
\be
{\mathcal Q}^{\rm (TT)}_{\ell m}(t) & =& R_p^{2+\ell} \int d\Omega\ Y^*_{\ell m}(\theta,\phi)
\nonumber \\ & \times &  \int_0^\infty dy 
\left( \frac{\rho(\vec{x},t)-\rho(y,\theta)}{\rho(y,\theta)} \right).
\label{eq:Qlm2}
\ee
Inspection of eq.\ref{eq:Qlm2} shows that the tide couples to an ``effective column"
\be
\Sigma(\theta,\phi,t) & \equiv  &
\int_0^\infty dy \left( \frac{ \rho(\vec{x},t)-\rho(y,\theta)}{\rho(y,\theta)}
\right).
\label{eq:Sigma}
\ee
Figure \ref{fig:colvst} shows numerical examples of the time-dependent column $\Sigma$ for 
the same simulations as in figures \ref{fig:TP_fullday_deep_Porb=4_Pspin=3} and
\ref{fig:TP_fullday_shallow_Porb=4_Pspin=3}.
The effective column (solid lines) in general leads the heating function, a consequence of
the temperature lag and the constant pressure assumption. We have taken a Fourier
transform of the time-series for $\Sigma$, and show the diurnal and semi-diurnal components, 
which also lead. The diurnal component dominates the temperature and
density response, but produces no torque since the tidal force has no $\ell=2$, $m=\pm 1$
components. The semi-diurnal response is smaller, by a factor $\sim 2$ in the present example,
but dominates the torque.

The expressions for the secular change in the planet's orbit require the Fourier 
transform in time of the quadrupole moments, which
we perform numerically from our simulation output. We define the desired transforms
\be
{\mathcal Q}^{\rm (TT)}_{\ell m}(t) & \equiv & \sum_{m=-\infty}^\infty 
{\mathcal Q}^{\rm (TT)}_{\ell m k} e^{-iknt}
\label{eq:Qlmexpand}
\ee
where
\be
{\mathcal Q}^{\rm (TT)}_{\ell m k} & = & \frac{n}{2\pi} \int_0^{2\pi/n} dt\ e^{iknt} 
{\mathcal Q}^{\rm (TT)}_{\ell m}(t),
\ee
and satisfy ${\mathcal Q}^{\rm (TT)}_{\ell ,-m,-k}=(-1)^m 
\left( {\mathcal Q}^{\rm (TT)}_{\ell m k}
\right)^\star$. For the nonlinear simulations, we evolved the atmosphere for 10 forcing periods,
and computed the Fourier transforms numerically using orbits 9 and 10.

We now turn to the gravitational tide.
The stellar gravitational tidal acceleration $-\grad U$ induces fluid flow in the planet.
Dissipation causes a phase lag $\sim 1/Q$ between the acceleration and high tide,
where $Q$ is the tidal quality factor.
This phase lag is the origin of the ``gravitational tide torque", which
attempts to make the planet's spin synchronous.
 \citet{1981A&A....99..126H}
contains a clear review of Darwin's theory of tidal evolution. There
it is assumed that the fluid response lags forcing by a time $\tau_{\rm
lag}$. The quadrupolar fluid response can be viewed as two point masses
each of mass $(k_p/2)M_\star[R_p/D(t-\tau_{\rm lag})]^3$, with longitude
$\Phi(t-\tau_{\rm lag})-\Omega(t-\tau_{\rm lag})$ and $\Phi(t-\tau_{\rm
lag})-\Omega(t-\tau_{\rm lag})+\pi$. Here $k_p$ is the apsidal motion
constant of the planet, which takes into account the internal structure
in the external potential perturbation produced by the quadrupole moment.
Under the assumption that $\tau_{\rm lag}$ is small in comparison to the 
forcing period, we expand to first order in $\tau_{\rm lag}$.
We ignore the term independent of 
$\tau_{\rm lag}$ as it is in phase with the stellar gravitational tidal 
forcing and therefore, cannot lead to a torque. 
We find the effective gravitational tidal quadrupole moment
\be
{\mathcal Q}^{\rm (GT)}_{2m}(t) & = & i \tau_{\rm lag} k_p M_\star R_p^2 Y_{2m}(\pi/2,0) 
\left( \frac{R_p}{a} \right)^3
\nonumber \\ && \times \sum_k (kn-m\Omega) X^{2m}_k(e) e^{-iknt},
\ee
which is defined in the inertial frame. 
Rewriting this expression in terms of $Q_p'=(3/2k_p)(1/n\tau_{\rm lag})$
\citep{2002ApJ...573..829M} and by taking the Fourier transform in time, 
we find the imaginary component
\be
{\rm Im} \left( {\mathcal Q}^{\rm (GT)}_{2 m k} \right) & = & \left( \frac{3}{2Q_p'} \right)
\left( \frac{M_\star R_p^5}{a^3} \right)
\left( \frac{kn-m\Omega}{n} \right)
\nonumber \\ && \times  Y_{2m}(\pi/2,0) X^{2m}_k(e).
\label{eq:hutQ}
\ee
As a check of this result, we compared eq.\ref{eq:hutQ}, \ref{eq:adotsec},
\ref{eq:edotsec} and \ref{eq:torquesec} against Hut's analytic formulas
for a range of orbital period, spin period and eccentricity.

Note that Hut's ``constant time lag" prescription for orbit and
spin evolution disagrees with the ``constant lag angle" approach of
\citet{1966Icar....5..375G}.  The use of a constant lag angle,
with a constant $Q$, is unphysical as the torque is discontinuous as
the spin changes from sub- to super-synchronous and vice-versa. 
By adopting the lag in time approach of 
\citet{1981A&A....99..126H} or \citet{2002ApJ...573..829M} with the
above choice for $Q$, we  find agreement with the results of 
\citet{1966Icar....5..375G} 
when the planet is far from a synchronous spin state, as in the 
case of the Jupiter-Io system.  Therefore, the constraints on $Q$
inferred by \citet{1966Icar....5..375G} may still be applied. Note, 
however, that
\citet{1978Natur.275...37I} and \citet{1980Icar...41....1D} utilize the
constant lag angle approach.   We find both qualitative and quantitative 
differences with their work.

\subsection{Linear Perturbation Theory}
\label{sec:linear}

For large forcing frequencies, the time-dependent temperature and flux changes
are small compared to the time-average values. One may then treat the 
time-dependent changes $\delta T(y,\theta,\phi,t)$ and $\delta F(y,\theta,\phi,t)$
as linear perturbations about the time-independent
background values $T(y,\theta)$ and $F(y,\theta)$.

\begin{figure}
\plotone{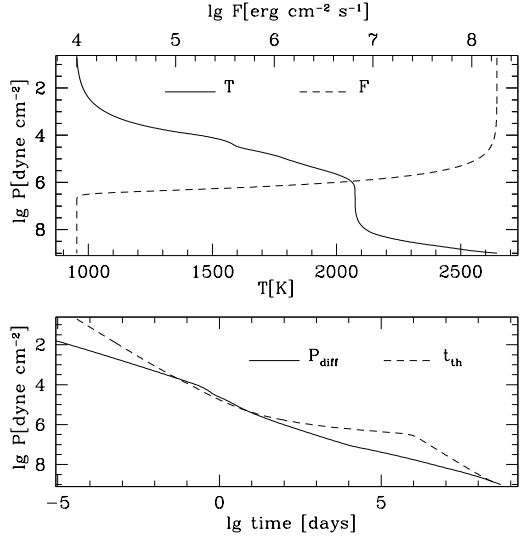}
\caption{ Background model constructed using time-averaged heating rate, 
using $\kappa_\star=10^{-2.5}{\rm cm^2\ g^{-1}}$, $F_{\rm core}=10^4\ {\rm erg\ cm^{-2}\
s^{-1}}$, $g=10^3\ {\rm cm\ s^{-2}}$, 
a solar-type star and $P_{\rm orb}=4\ {\rm days}$.}
\label{fig:bg_kapa=0.003_porb=4.0day}
\end{figure}

The background model is computed by integrating
\be
\frac{dT}{dy} & = & \frac{\kappa F}{16 \sigma_{\rm sb} T^3\Lambda}
\label{eq:bgdTdy}
\ee
with flux found by integrating eq.\ref{eq:epsbg}
\be
&& F(y,\theta)   =   F_{\rm core} + \int_y^\infty dy'\ \epsilon(y',\theta)
\nonumber \\ & = & 
F_{\rm core}  +  
\frac{F_\star(a)\sin\theta X^{10}_0(e)}{\pi} \int_0^{\pi/2} d\phi
\cos\phi\ e^{-(\kappa_\star y/\sin\theta) \sec\phi},
\label{eq:bgF}
\ee
and subject to the boundary condition
$T\simeq (F(0,\theta)/4\sigma_{\rm sb})^{1/4}$ at small optical depth.
We construct this model by integrating inward from the surface.
An example is given in figure \ref{fig:bg_kapa=0.003_porb=4.0day}. 
Our treatment of radiative transfer is simplified, but nevertheless reproduces
the main features of full solutions to the transfer equations. At small
optical depth the solution becomes isothermal at the ``skin" temperature.
Heating by inward-going stellar photons, here through the   heating 
function, generates an equal outward flux of
thermal photons (for the
time-independent case). At a column $y \simeq\kappa^{-1}_{\star}$, the heating function
and outward flux decrease exponentially, and the temperature profile becomes 
isothermal at $T(\kappa_\star^{-1},\theta) \simeq
(F(0,\theta)/\sigma_{\rm sb})^{1/4}(\kappa/\kappa_\star)^{1/4}$. 
Lastly, the flux $F_{\rm core}$ from deep in the atmosphere eventually
causes the temperature to increase inward, the gradient becoming steep enough to
cause convection.

The linearized equations are found by perturbing
eq.\ref{eq:fullheat} and \ref{eq:fullflux} while enforcing strict hydrostatic balance
for the perturbations, so that each fluid element remains at constant $y$. 
We eliminate dependence on $\phi$ and $t$ by expanding $\delta T$ and $\delta F$ 
in in a Fourier series (e.g.,
eq.\ref{eq:epsmk}). The time dependence of the perturbations is then
$d/dt = - i \sigma_{mk}$. We will
suppress $m$ and $k$ in the following equations. The 
linearized form of eq.\ref{eq:fullheat} and \ref{eq:fullflux} are 
\be
\frac{\partial \delta T}{\partial y} & = & \frac{dT/dy}{1-\Lambda_R} 
\left[ \frac{\delta F}{F}
+ \left\{ \kappa_T - 3 - \Lambda_R(\kappa_T+1)\right\} \frac{\delta T}{T} \right]
\label{eq:Tpert}
\ee
and 
\be
\frac{d \delta F}{d y} & = &  -i\sigma C_p \delta T - \delta \epsilon,
\label{eq:Fpert}
\ee
where $\kappa_T=\partial \ln \kappa/\partial \ln T |_P$
and
\be
\Lambda_R & = & \frac{R^2(R+4)}{(6+3R+R^2)(2+R)}.
\ee
At large (small) optical depth, $\Lambda_R \rightarrow 0\ (1)$.
We solve eq.\ref{eq:Tpert} and \ref{eq:Fpert} subject to the boundary conditions
$\delta F/F=4\delta T/T$ at the top of the grid and 
$\delta T, \delta F \rightarrow 0$ at the base.
The equations are solved by finite difference in $y$ to
obtain a banded matrix equation which is readily inverted to find
the response $\delta T$ and $\delta F$ to the heating $\delta \epsilon$.
The top of the grid is set at a column $10^{-2}$ times smaller than the thermal
photosphere, and the base of the grid is set at the radiative-convective boundary,
typically well below $y=\kappa_\star^{-1}$.

\begin{figure}
\plotone{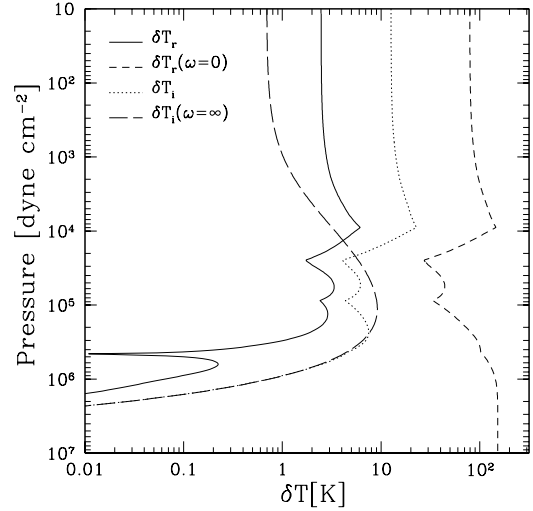}
\caption{ Semi-diurnal ($m=2$), linear temperature perturbations for a planet at $P_{\rm orb}=4$ days
with a prograde rotation rate of $P_{\rm spin}=1.0$ day, giving a semi-diurnal
forcing period of $2/3$ day. Surface gravity is $g=10^3\ {\rm cm\ s^{-2}}$,
$\theta=\pi/2$, and $\kappa_\star=10^{-2.5}\ {\rm cm^2\ g^{-1}}$. The high frequency limit
from eq.\ref{eq:dThighfreq} is a good approximation for $P \ga 10^4 {\rm dyne\ cm^{-2}}$. }
\label{fig:linear_deep_highfreq}
\end{figure}

\begin{figure}
\plotone{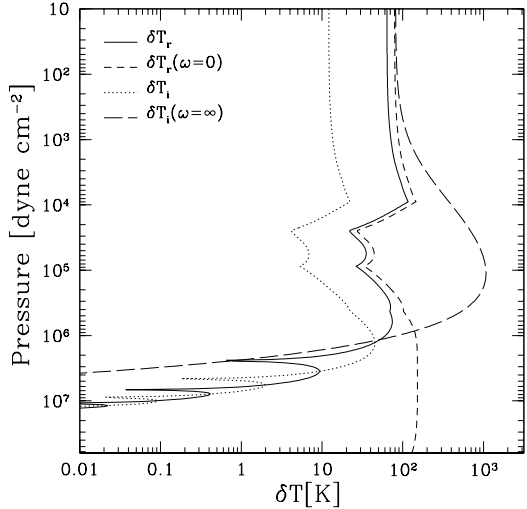}
\caption{ Same as figure \ref{fig:linear_deep_highfreq} but with a
spin period $P_{\rm spin}=3\ {\rm days}$
and shallow heat deposition
$\kappa_\star=10^{-0.5}\ {\rm cm^2\ g^{-1}}$. The low frequency limit is a good
approximation for $P \leq 10^6 {\rm dyne\ cm^{-2}}$, below which the oscillations 
in the real and imaginary part are evidence of a diffusion wave. }
\label{fig:linear_shallow_lowfreq}
\end{figure}

Given the temperature perturbation, the density perturbation is
$\delta \rho/\rho = - \delta T/T$ at constant pressure. To linear order, 
the effective column is given by 
\be
\Sigma(\theta) & = & \int_0^\infty dy 
\left( -
\frac{\delta T(y,\theta)}{T(y,\theta)} \right),
\label{eq:sigmalin}
\ee
implying that the temperature perturbations {\it lagging} the heating
tend to torque the planet away from synchronous rotation.

Examples of the solution of the linearized equations are shown in figures
\ref{fig:linear_deep_highfreq} and \ref{fig:linear_shallow_lowfreq},
along with analytic approximations to be discussed presently. The
difference between these two plots is mainly due to the depth of heating, i.e.
deep in figure \ref{fig:linear_deep_highfreq} and shallow in figure
\ref{fig:linear_shallow_lowfreq}. 

For the deep heating case in figure \ref{fig:linear_deep_highfreq},
the forcing period ($=2/3$ day) was chosen to be shorter than the
diffusion time for the base of the heated layer.
There is no time for diffusion to transport heat,
the $\partial F/\partial y$ term in eq.\ref{eq:fullheat} is small,
and the temperature fluctuates solely due to the time-dependent heating.
A longer forcing period leads to larger input of heat, and effective 
column $\Sigma \propto P_f$, where $P_f=2\pi/|\sigma|$ is the forcing period.
An analytic solution will be
presented in \S \ref{sec:highfreq}. Since the imaginary part of the
temperature perturbation is large only above $P_{\rm base}=g/\kappa_\star$,
the torques will extend down to $P_{\rm base}$.

The shallow case in figure \ref{fig:linear_shallow_lowfreq}, on the
other hand, has a sufficiently long forcing period ($6$ days) that the
perturbations extend deeper than $P_{\rm base}$ by some 2 orders of
magnitude in pressure. For pressures $P \la 10^6\ {\rm dyne\ cm^{-2}}$,
the first term in eq.\ref{eq:fullheat} can be ignored due to the small
forcing frequency. In this limit, the atmosphere responds in phase
to the forcing, changing quasi-statically from one equilibrium to
another. No torque is exerted on the atmosphere in this limit since the
temperature perturbation is in phase with the heating
function. That is, for the semi-diurnal mode, temperature maxima occur 
at noon and midnight, while temperature minima take place at sunrise and 
sunset.  In this ``$\sigma=0$" limit, we can solve for the (real)
temperature perturbation by integrating eq.\ref{eq:fullheat} with $d
T/d t=0$.  In this regime, the thermal inertia of the material is small 
is comparison to the amount of photon energy absorbed per cycle and therefore, 
the temperature is ``locked'' to the heating function. 
For the forcing frequencies of interest, this limit will
always be invalid sufficiently deep in the envelope, at which point 
the solution is well described by a diffusion wave.
This wave is apparent in figure
\ref{fig:linear_shallow_lowfreq} in the region $P=10^6-10^7{\rm dyne\
cm^{-2}}$, where the real and imaginary parts are comparable, and exhibit
oscillations. The first and second terms in eq. \ref{eq:fullheat} are
comparable for diffusion waves below the base of the heating layer.
The decay of the envelope of this wave into the planet determines the column
to which torque is applied. 

Using dimensional analysis on eq.\ref{eq:fullheat} and \ref{eq:fullflux}, 
heat can diffuse down to a column
\be
y_{\rm diff}  &=&  \sqrt{ \frac{16\sigma_{\rm sb} T^3}{3\kappa C_p |\sigma|} }
\ee
in a forcing period $P_f=2\pi/|\sigma|$. This depth is to 
be compared to the base of the
heating layer at $\kappa_\star^{-1}$. Heating can diffuse below 
the base of the heating layer when
\be
P_f & \ga &  P_{\rm diff}  =   \frac{3\pi}{8} \frac{\kappa C_p}{\sigma_{\rm sb} 
T^3 \kappa^2_{\star}}\nonumber\\
& = & 3.6\ {\rm days}\ \left(\frac{\kappa}{0.1\ {\rm cm^2\ g^{-1}}}\right)
\nonumber\\
& \times & 
\left(\frac{0.01 \ {\rm cm^2\ g^{-1}}}{\kappa_\star} \right)^2
\left( \frac{2000\ {\rm K}}{T} \right)^3.
\label{eq:Pdiff}
\ee
In the low frequency limit, the effective column
\be
\Sigma & \sim & y_{\rm diff} \propto P_f^{1/2},
\ee
which increases more slowly as $P_f \rightarrow 0$ than the high frequency
limit. 

Close to the star, where the gravitational tide is large,
the equilibrium spin must be in the long forcing period limit, and far
from the star, where the gravitational torque is small, the equilibrium
spin must be in the short forcing period limit.  The dividing line
between these two limits depends on factors such as stellar flux,
opacities and tidal $Q$.

Unless $\partial T/\partial y = T/y$, the ``diffusion depth",
$y_{\rm diff}$, differs from the often-used ``cooling depth",
$y_{\rm cool}=F/(C_pT|\sigma|)$. See the lower panel in figure
\ref{fig:bg_kapa=0.003_porb=4.0day} for a numerical example. In the
problem at hand, $y_{\rm cool}$ and $y_{\rm diff}$ can differ by orders
of magnitude, especially at small optical depths and below the base
of the heating layer where the flux drops exponentially. Physically,
$y_{\rm cool}$ is the depth down to which the heat content can be
radiated in the timescale $2\pi/|\sigma |$. In the absence of heating,
the flux tends to become constant with depth for $y \la y_{\rm cool}$
(e.g. \citealt{2005ApJ...628..401P}). As we will show in the next
sections, $y_{\rm diff}$ is the depth down to which small perturbations
can propagate, and hence is the more relevant for heating deep in the
atmosphere. 


\subsection{ Analytic solution in the high frequency limit: $\sigma\,t_{\rm th}\gg 1$}
\label{sec:highfreq}

The thermal time at the base of the absorbing layer, 
$t_{\rm th} \sim C_p T/\kappa_\star F_\star$, is the timescale
over which the layer can heat or cool. For large forcing frequencies
$\sigma t_{\rm th} \gg 1$, heat cannot diffuse over significant
distances in a forcing period and the term $d\delta F/dy$ can be ignored. 
In this limit, the temperature perturbation is determined solely by 
the local heating function (eq.\ref{eq:Fpert}), i.e., 
\be
\delta T_{mk} & = & i \frac{\delta \epsilon_{mk}}{\sigma_{mk} C_p},
\label{eq:dThighfreq}
\ee
where $\delta \epsilon_{mk}$ is the appropriate Fourier coefficient in eq.\ref{eq:epsmk}.
This expression shows that the temperature perturbations are small in the high frequency
limit, $\delta T$ decreases exponentially below $y = \kappa_\star^{-1}$
and its magnitude increases with forcing period since there is more time
to absorb heat. 
With respect to the forcing frequency $\sigma_{mk}$, the temperature 
perturbation lags the forcing by $90^\circ$ in phase. 
It is compared to the solution of the linearized
boundary value problem in figure \ref{fig:linear_deep_highfreq}.

The effective column is then
\be
\Sigma_{mk}(\theta) & \simeq & - \frac{i}{\sigma_{mk} C_p} \int_0^\infty 
dy \left( \frac{\delta
\epsilon_{mk}(y,\theta)}{T(y,\theta)} \right).
\label{eq:Sigmamk}
\ee
Since $T(y,\theta)$ varies relatively slowly for $y \la \kappa_\star^{-1}$, we
approximate it as a constant 
and pull it out of
the integral. 
The remaining integral is
\be
\int_0^\infty ds\ g_m(s)  & \equiv & G_m,
\label{eq:Gm}
\ee
where $G_0=1/\pi$ and $G_2=1/3\pi$. Plugging eq.\ref{eq:Gm} into eq.\ref{eq:Sigmamk} we find
\be
\Sigma_{mk}(\theta) & \simeq & - i \frac{F_\star(a)X^{1m}_k(e)G_m\sin\theta}{\sigma_{mk} C_p T}.
\label{eq:Sigmamkhighfreq}
\ee
For the diurnal (semi-diurnal) component, eq.\ref{eq:Sigmamkhighfreq} implies $\Sigma$ leads the 
heating function by $\pi/2$ ($\pi/4$) in longitude. This phase lead corresponds to 
6am (9am) for the maxima of the diurnal (semi-diurnal) components in figure \ref{fig:colvst}.

The quadrupole moments require the angular integrals
\be
F_{\ell m} & = & 2\pi \int_0^\pi d\theta \sin^2\theta Y_{\ell m}(\theta,0).
\ee
For example, $F_{22}=\sqrt{15\pi/2}(3\pi/16)$ and $F_{20}=(-\pi/16)\sqrt{5\pi}$.
The final result for the quadrupole moments in the high frequency limit is
\be
{\rm Im} \left( {\mathcal Q}^{{\rm (TT)}}_{\ell m k} \right)
& = & - \frac{R_p^{2+\ell}F_\star(a)}{\sigma_{km} C_p T}  X^{1m}_k(e) F_{\ell m} G_m,
\label{eq:TTQlm_highfreq}
\ee
confirming the order of magnitude estimates found in eq.\ref{eq:TTQest} and \ref{eq:tth}
if we set $\Delta M=R_p^2/\kappa_\star $, $t_{\rm th}=C_p T /\kappa_\star F_\star(a)$
and $\delta=1$.

Eq.\ref{eq:TTQlm_highfreq} agrees with the quadrupole moment found in
\citet{1978Natur.275...37I}, up to the uncertainty as to what value of
temperature should be used when it is pulled out of the integral. These
authors derived the high frequency limit appropriate for a circular orbit
and asynchronous rotation, and implicitly assumed a finite atmosphere
bounded below by a hard surface. Eq.\ref{eq:TTQlm_highfreq} generalizes
their result to a deep atmosphere, and allows for an eccentric orbit.
 

\subsection{ Analytic solution in the low frequency limit: 
$\sigma\, t_{\rm th}\ll 1$ }
\label{sec:lowfreq}

We now consider the low frequency limit $\sigma t_{\rm th} \ll 1$, in which
the thermal time in the layer directly heated by insolation ($y \la \kappa_\star^{-1}$) is short
compared to the forcing period. The small thermal inertia of the
absorbing layer implies that temperature perturbations
are large and in phase with the forcing, with corresponding 
density perturbations $180^\circ$
out of phase with the insolation, implying that the torque is zero.
There will always be a layer at $y \ga
\kappa_\star^{-1}$ sufficiently deep that the thermal time becomes longer
than the forcing period, and the temperature perturbation in this deep
layer will be out of phase, yielding a torque.

\begin{figure}
\plotone{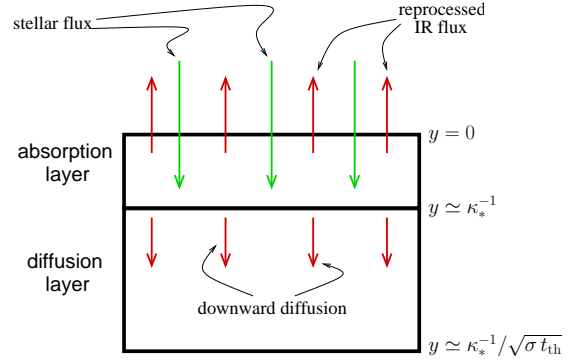}
\caption{ Cartoon depicting time-dependent heating in the limit where
the forcing  period, $2\pi/\sigma $, is large in comparison to the thermal time
$t_{\rm th}$ of the absorption layer.  The temperature of the absorption layer
is approximately ``locked'' to the time-varying insolation such that the maxima in
temperature take place at noon for diurnal heating and noon and midnight for
semi-diurnal heating.  However, a thermal diffusion wave penetrates
to a column depth $y\sim y_{\rm diff}=\kappa^{-1}_*/\sqrt{\sigma\,t_{\rm th}}$.
As expected for a thermal diffusion wave, the temperature fluctuation at $y\sim y_{\rm diff}$
lags the temperature fluctuation in the absorption layer by $\pi/4$ in phase,
which ultimately allows the heated layers to couple to the semi-diurnal
tidal field in such a way as to increase its spin angular momentum.
}
\label{fig: two_layer}
\end{figure}

To illustrate this, we divide the atmosphere into two separate plane-parallel layers, as
depicted in Figure \ref{fig: two_layer}.  The stellar flux is absorbed
in the upper layer at a column $y \sim \kappa_\star^{-1}$, and generates
an outward flux of thermal radiation. In the lower layer, the stellar
flux is negligible, and time-dependent temperature perturbations can only
penetrate this layer due to thermal diffusion.

The thermal time in the upper layer is short, and the time derivative term in eq.
\ref{eq:Fpert} can be set to zero. The temperature profile changes quasi-statically
in response to changes in the heating rate. As an analytic example, consider
a flux profile $F(y)=F_0\exp(-\kappa_\star y)$ and constant opacities $\kappa$ and 
$\kappa_\star \ll \kappa$. The temperature deep in the layer is then
$T \simeq (3\kappa F_0/4\sigma_{\rm sb} \kappa_\star)^{1/4}$. If the heating produces
a change in flux $F_0 \rightarrow F_0 + \delta F_0$, the fractional change in temperature
deep in the upper layer is 
\be
\frac{ \delta T}{T}(y=\kappa_\star^{-1}) & \simeq & \frac{\delta F_0}{4 F_0}.
\label{e: deltaTbndy}
\ee
For semi-diurnal 
forcing, the fact that $\delta T$ reaches it's maximum at
noon and midnight implies that $\delta\rho$, or the thermal tidal 
bulge, achieves it's maximum value at sunrise and sunset.  As result,
the gravitational tidal field cannot apply a torque to the 
the thermal tidal bulge induced in the upper layer in the $\sigma t_{\rm th} \ll 1$
limit.

Direct heating is negligible in the lower layer, and $\delta \epsilon$ in 
eq. \ref{eq:Fpert} may be set to zero.  Therefore, the temperature 
perturbation is determined by a source-free diffusion equation 
subject to the boundary condition in eq.\ref{e: deltaTbndy} at $y=\kappa^{-1}_\star$.
The background temperature $T$ is roughly 
constant below $y=\kappa^{-1}_\star$, and we assume constant opacity $\kappa$ for simplicity.
With this, the temperature perturbation in the lower diffusion layer
obeys 
\be
-i\sigma C_p \delta T=\frac{16\sigma_{\rm sb}^3T^3}{3\kappa}\frac{d^2\delta T}{
dy^2}
\label{eq:diffeqn}
\ee   
over the interval $\kappa_\star^{-1} \leq y \leq \infty$.
The general solution of eq.\ref{eq:diffeqn} has the form 
\be
\frac{\delta T}{T}= Ce^{s(y-\kappa_\star^{-1})}+De^{-s(y-\kappa_\star^{-1})}
\ee
where 
\be
s=e^{-i\pi/4}\sqrt{\frac{3\kappa \sigma C_p}{
16\sigma_{\rm sb} T^3}}
\ee
is the wavenumber, in units of inverse column. Note that $\sigma$ may be either 
positive or negative in this expression. Finiteness of the solution at $y=\infty$
implies $C=0$ for either sign of $\sigma$. 
The final result for the temperature perturbation in the lower layer is then
\be
\frac{\delta T}{T}(y) = \frac{\delta F_0}{4 F_0} e^{-s(y-\kappa_\star^{-1})}.
\label{eq:dTlower}
\ee
Since $s$ is complex, eq.\ref{eq:dTlower} implies
oscillations in temperature, as well as an envelope decreasing to larger
depths. These oscillations are apparent in the linear solution in figure
\ref{fig:linear_shallow_lowfreq}, as well as in the nonlinear solution
in figure \ref{fig:TP_fullday_deep_Porb=4_Pspin=3}.

The effective column is determined by integrating eq.\ref{eq:dTlower} over the 
lower layer to find
\be
\Sigma & = & -\int_{\kappa_\star^{-1}}^\infty dy\ \frac{\delta T}{T} 
\simeq - \left( \frac{\delta F_0}{4 F_0} \right) \left( \frac{1}{s} \right)
\nonumber \\
& = & - e^{i\pi/4} \left( \frac{\delta F_0}{4 F_0} \right) \sqrt\frac{16\sigma_{\rm sb} T^3}{3\kappa \sigma C_p}
\label{eq:sigmalowfreq}
\ee
which leads in phase by $3\pi/4$, as compared to $\pi/2$ for the high frequency limit.
For the diurnal (semi-diurnal) component, eq.\ref{eq:sigmalowfreq} implies $\Sigma$ leads the
heating function by $3\pi/4$ ($3\pi/8$) in longitude. This phase lead corresponds to
3am (7:30am) for the maxima of the diurnal (semi-diurnal) components in figure \ref{fig:colvst}.

Combining the results from \S \ref{sec:highfreq} and this section, we find generically
that the semi-diurnal component of $\Sigma$, and hence the quadrupole moments, have
the {\it opposite} phase as the gravitational tide quadrupole moments.
Hence they promote asynchronous spin, and may drive eccentricity.
Furthermore, eq.\ref{eq:sigmalowfreq} shows that $\Sigma$ continues
to increase as the forcing frequency goes to zero, albeit with a shallower power of $\sigma$. 
We summarize the scalings from eq.\ref{eq:Sigmamkhighfreq} 
and \ref{eq:sigmalowfreq} by writing
\be
\Sigma\sim\kappa^{-1}_{\star}\,{\rm Min}\,[ (\sigma\,t_{\rm th})^{-1},
(\sigma\,t_{\rm th})^{-1/2}],  
\label{eq:fit}
\ee
and phase lead in the range $\pi/2-3\pi/4$.


\section{ Equilibrium spin rate }
\label{sec:spineq}

Hot Jupiters are commonly assumed to be very nearly synchronized. The argument
is that the synchronization timescale, for $Q_p'=10^6$,
is far shorter than the age of observed planets.
The torque on the planet due to the gravitational tide is given by (eq.\ref{eq:hutQ}
and \ref{eq:torquesec})
\be
N^{{\rm (GT)}} & = & \left( \frac{9}{2} \right) \left( \frac{n}{Q_p'} \right)
\left( \frac{M_\star^2}{M_\star+M_p} \right)
\left( \frac{R_p^5}{a^3} \right) \left( n-\Omega \right)
\nonumber \\ & = & 
- 7 \times 10^{32}\ {\rm erg}\  R_{10}^5 \left( \frac{10^6}{Q_p'}\right)
 \left( \frac{P_{\rm orb}}{4\ {\rm days}} \right)^{-4}
\left( \frac{\Omega-n}{n} \right)
\label{eq:NGTcirc}
\ee
for a circular orbit. Starting from a rapid initial spin rate $\Omega_0 \gg n$,
the synchronization time is
\be
t_{\rm synch, GT} & = & - \frac{I_p \Omega_0}{N^{\rm (GT)}}
= \frac{2}{9} \eta \left( \frac{Q_p'}{n} \right) \left( \frac{M_p(M_\star+M_p)}{M_\star^2} \right)
\left( \frac{a^3}{R_p^3} \right)
\nonumber \\ & = & 
10^5\ {\rm yr}\ \left( \frac{\eta}{0.25} \right) \left( \frac{Q_p'}{10^6} \right)
\left( \frac{2.8\ {\rm hr}}{P_p} \right)^2 \left( \frac{P_{\rm orb}}{4\ {\rm days}} 
\right)^3.
\ee
where $I_p=\eta M_p R_p^2$, $\eta \simeq 0.25$ is appropriate for 
Jupiter-like planets, and we have defined the ``dynamical time" $P_p=2\pi
(R_p^3/GM_p)^{1/2}$. Hence, for $Q_p'=10^6$, planets out to $P_{\rm orb} \sim
150\ {\rm days}$ should be synchronized in $5\ {\rm Gyr}$.

The fact that gas giants have $Q_p' \sim 10^6$ implies gravitational
tides are far {\it weaker} than for terrestrial type planets with $Q_p' \sim 10-100$.
Additional torques are relatively more important for gas giants in comparison to terrestrial planets.
We now show that thermal tidal torques are capable of generating significant 
asynchronous spin for hot Jupiters.

The high frequency limit has a simple analytic solution. For zero
eccentricity, plugging the $\ell=m=k=2$ quadrupole moments from
eq.\ref{eq:TTQlm_highfreq} into eq.\ref{eq:Ntt2} gives the thermal tide torque
in the high frequency limit
\be
N^{(\rm TT)} & = & \frac{3\pi}{16} \frac{n^2 R_p^4 F_\star}{(\Omega-n) C_p
T }.
\ee
Recall that $T$ should be evaluated at $y \sim \kappa_\star^{-1}$, and here $F_\star=F_\star(a)$.
This estimate agrees with that of \citet{1978Natur.275...37I}. Plugging in
numbers for a solar-type star we find
\be
N^{(\rm TT)} & = & 2.8 \times 10^{32}\ {\rm erg} \left( \frac{n}{\Omega-n} \right)
\left( \frac{4\ {\rm days}}{P_{\rm orb}} \right)^2 R_{10}^4 f_T^{-1}
\label{eq:Ntt3}
\ee
where $R_{10}=R/10^{10}\ {\rm cm}$ and $f_T=T/(F_\star/\sigma_{\rm sb})^{1/4}$. 
Clearly for the hot Jupiters the gravitational tide torque in eq.\ref{eq:NGTcirc}
and the thermal tide torque in eq.\ref{eq:Ntt3} are comparable.
Equating eq.\ref{eq:Ntt3} and \ref{eq:NGTcirc} we find the
equilibrium spin rate
\be
\Omega_{\rm eq} & = & n \left( 1 \mp \sqrt{
\frac{\pi}{24} \frac{F_\star}{C_pT} \frac{GQ_p'}{n^3 R_p} } \right)
\label{eq:omegaeq}
\ee
or, numerically
\be
P_{\rm spin,eq} & = & \frac{P_{\rm orb}}{ \left( 1 \pm 
0.63\ \left( \frac{P_{\rm orb}}{4\ {\rm days}}
\right) \left( \frac{Q_p'}{10^6} R_{10}^{-1} f_T^{-1} \right)^{1/2} \right)}.
\label{eq:asynch}
\ee
The $\mp$ sign denotes the subsynchronous ($\Omega<n$) and supersynchronous
($\Omega>n$) solutions. The subsynchronous solution becomes retrograde
($\Omega<0$) for sufficiently long $P_{\rm orb}$. 
Eq.\ref{eq:asynch} shows that the degree of asynchronous spin increases for larger
$P_{\rm orb}$ and is large for the hot Jupiters for $Q \sim 10^5-10^6$.
At long orbital period, the spin period asymptotes to a constant
\be
P_{\rm spin,eq} & \longrightarrow & \pm 6.34\ {\rm days} 
\left( R_{10} \frac{Q_p'}{10^6}  f_T^{-1} \right)^{1/2}.
\ee
If no new physics intervenes, this asymptotic spin rate should hold for planets
at $P_{\rm orb} \la 150$ days which have sufficient time to attain spin equilibrium.
Since the degree of asynchronous spin decreases toward the star, the high frequency
limit becomes less accurate there. 

Small perturbations about the equilibrium spin rate, at fixed $P_{\rm orb}$ and $R_p$,
are stable for both the super- and sub-synchronous solutions given above.


The accuracy of the high frequency limit can be estimated analytically. The semi-diurnal ($m=2$) forcing period in the high frequency limit is
\be
P_f & = & 3.2\ {\rm days}\ R_{10}^{1/2} \left( \frac{10^6}{Q_p'} \right)^{1/2},
\label{eq:Pf2}
\ee
independent of orbital period. The diffusion time to the base of the heating layer
is (eq.\ref{eq:Pdiff})
\be
P_{\rm diff} & \simeq & 1.6\ {\rm days}\ \left( \frac{P_{\rm orb}}{4\ {\rm days}} \right)
\left( \frac{ 10^{-2.5}\ {\rm cm^2\ g^{-1}} }{\kappa_\star} \right).
\label{eq:Pdiff2}
\ee
At long orbital periods, the forcing period becomes shorter than the diffusion time, and
the high frequency limit is applicable. Equating the expressions in eq.\ref{eq:Pf2} and
\ref{eq:Pdiff2} a rough estimate for the applicability of the high frequency limit is
\be
P_{\rm orb} & \geq &  8\ {\rm days}\ R_{10}^{1/2} \left( \frac{10^6}{Q_p'} \right)^{1/2}
\left( \frac{\kappa_\star}{ 10^{-2.5}\ {\rm cm^2\ g^{-1}}  } \right).
\ee

\begin{figure}
\plotone{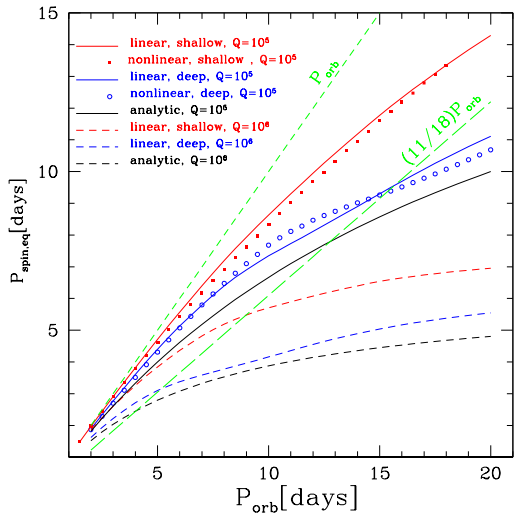}
\caption{ Equilibrium spin period as a function of orbital period for super-synchronous solutions
($\Omega>n$). The ``shallow" and ``deep"
lines represent $\kappa_\star=10^{-0.5},10^{-2.5}\ {\rm cm^2\ g^{-1}}$,
respectively. The solid black line is the analytic model from \S
\ref{sec:highfreq} with $f_T=1$. The short-dashed green line is
the orbital period, and the long-dashed green line shows $P_{\rm spin}=(11/18)P_{\rm orb}$, the
spin period below which the gravitational tide excites eccentricity.
Surface gravity is $g=10^3\ {\rm cm\ s^{-2}}$ and radius is $R_p=10^{10}\ {\rm cm}$. The 
flux from the core is taken to be $F_{\rm core}=10^4\ {\rm erg\ cm^{-2}\ s^{-1}}$, independent
of latitude. }
\label{fig:eqspin}
\end{figure}

The high frequency limit allows useful analytic intuition, but neglects
diffusion and approximates temperature at the base as constant. We
now present results on equilibrium spin rate and heating rate relaxing
these assumptions.
Note, however, that these calculations are not fully self-consistent in
that we fix a radius, and we also take, for simplicity, the flux from the core
to be constant, independent of latitude. In \S \ref{sec:thermal}
we will self-consistently determine the radius in thermal equilibrium and determine
the self-consistent value of $F_{\rm core}$.

 Figure \ref{fig:eqspin} shows equilibrium spin rate
for three different methods of calculation: ``nonlinear simulations"
as discussed in \S \ref{sec:eqgeom} and \ref{sec:tprofile},
``linear" calculations solving the linearized equations as a boundary
value problem (\S \ref{sec:linear}), and the ``analytic" result
from eq.\ref{eq:asynch} in the high frequency limit. Two different heating
depths are shown, ``shallow" and ``deep" lines, as well as two different
values for the tidal $Q$. 
The good agreement between linear and nonlinear calculations
implies that at the large depths where the quadrupole moments are determined, nonlinear
effects do not greatly change the result. 
The analytic approximation is accurate to with $\sim 10\%$.
The degree of asynchronous spin is large for both $Q_p'=10^5$ and $10^6$ cases.
The shallow heating case is more nearly synchronized than the deep heating case as it is 
more in the low frequency limit than the deep heating case, and hence the thermal tide 
torque is weaker.



\section{Tidal heating rates}
\label{sec:radii}


\begin{figure}
\plotone{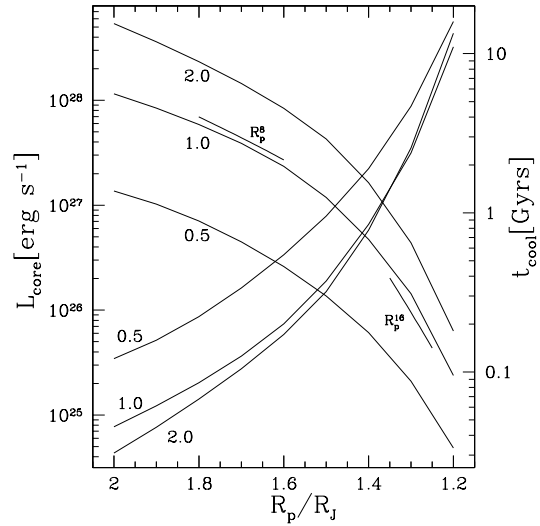}
\caption{ Cooling luminosity from the core (curves sloping up to the left) and cooling time (curves
sloping up to the right) versus 
planetary radius for three different planetary masses, $M_p/M_J=0.5, 1.0, 2.0$.
A solar type star is used and the planets sit at $P_{\rm orb}=2$ days. The heating depth
is given by $\kappa_\star=10^{-2}\ {\rm cm^2\ g^{-1}}$. Closer to the star, insolation is larger 
and $L_{\rm core}$ decreases, and visa versa further from the star.}
\label{fig:lum_vs_rad}
\end{figure}

The large tidal heating rates for eccentric hot Jupiters can inflate
the radii of these planets if heat is deposited in the convective core,
or slow their cooling if heat is deposited in the radiative envelope
\citep{2001ApJ...548..466B}. Previous investigations considered
transient tidal heating due to an energy reservoir in the initial
eccentricity, or steady-state heating due to eccentricity pumped by a
companion. In this section we compute gravitational tidal heating rates due to 
the asynchronous spin set by a balance of gravitational and thermal tide
torques for a planet on a circular orbit. We compare these heating rates
to the cooling rate of the core to determine if tidal heating can power the large
observed radii.
 

Appendix \ref{sec:appendix2} describes an updated version of the planetary
structure and evolution models of \citet{2006ApJ...650..394A}.
Figure \ref{fig:lum_vs_rad} shows the cooling luminosity of the convective
core as a function of radius for $P_{\rm orb}=2\ {\rm days}$ around
a Sun-like star. Also shown is the timescale to attain thermal equilibrium
\be
t_{\rm cool} & = & \frac{\int_0^{M_p} dm\ C_p T}{L_{\rm core}}.
\label{eq:tcool}
\ee
Cooling timescales for Jupiter-mass planets become shorter than a typical age of $3\ $ Gyr for 
$R_p \ga (1.3-1.4) \times R_J$. Cooling rates closer (further) from the
star are lower (higher) due to the insulation effect.

Models for the passive cooling of irradiated hot Jupiters can typically
explain planets with radii $R \la 1.2R_J$ and below, but have difficulty
slowing the cooling enough to explain planets with $R \ga 1.2R_J$
(e.g. \citealt{2008arXiv0801.4943F}).  If we are interested in the radius
range $R=(1.2-1.8) \times R_J$, figure \ref{fig:lum_vs_rad} shows this corresponds to
core cooling rates of $10^{25}-10^{28}\ {\rm erg\ s^{-1}}$, depending on
mass and insolation level. In order for tides to halt the cooling of
the core, this loss of heat at the radiative-convective boundary must be
balanced by a gain of heat in the core due to tides. 

The tidal heating rate in eq.\ref{eq:heatrate} is related to the torque in eq.\ref{eq:torquesec}
by the
pattern speed $(kn-m\Omega)/m$ for each harmonic. In spin equilibrium, with
$N^{\rm (TT)}=-N^{\rm (GT)}$, the gravitational tide heating rate for the high frequency limit is
\be
\dot{E}^{\rm (GT)}_{\rm heat} & = & (n-\Omega) N^{\rm (GT)}
\nonumber \\ & = &
  2.9\times 10^{27}\ {\rm erg\ s^{-1}}\
\left( \frac{4\ {\rm days}}{P_{\rm orb}} \right)^3 R_{10}^4.
\label{eq:Edotgt}
\ee
The tidal heating rate in the high frequency limit is independent of $Q$
since $\Omega-n \propto (Q_p')^{1/2}$ and $\dot{E} \propto (\Omega-n)^2/Q_p'$.

\begin{figure}
\plotone{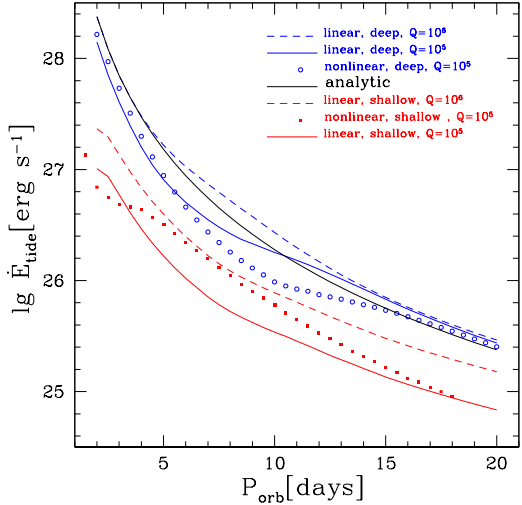}
\caption{ Gravitational tide heating rate for the equilibrium spin periods shown in figure \ref{fig:eqspin}. }
\label{fig:eqheat}
\end{figure}

Figure \ref{fig:eqheat} shows the tidal heating rates for the equilibrium spin rates
in figure \ref{fig:eqspin}. The same small deviations between the linear and nonlinear
results are apparent in this figure, but note that for the deep heating case the nonlinear,
linear and analytic (high frequency limit) results agree at large orbital period. The
shallow heating case shows a smaller heating rate than the deep heating case as the torque
is smaller. Recall that the tidal heating rate is independent of $Q_p'$ in the high frequency
limit. This agrees with the deep heating case for large $P_{\rm orb}$.
Somewhat counterintuitively, {\it larger } $Q_p'$ leads to a larger tidal heating rate at smaller
$P_{\rm orb}$ when thermal diffusion effects become important. The dependence of spin and
heating rates on $Q_p'$ can be understood simply.
Let $N^{\rm (TT)} = a(n)(\Omega-n)^{-\alpha}$
and $N^{\rm (GT)}=-b(n) (\Omega-n)^\beta Q^{-1}$, for some functions $a$ and $b$ of orbital period,
and exponents $\alpha$ and $\beta$. The equilibrium spin rate is
\be
\Omega_{\rm eq}-n & = & \left( \frac{aQ}{b} \right)^{1/(\alpha+\beta)}
\ee
and the tidal heating rate is
\be
\dot{E}_{\rm G} & = & (\Omega-n)^{1+\beta} b Q^{-1} = a^{(\beta+1)/(\beta+\alpha)}
(Q/b)^{(1-\alpha)/(\beta+\alpha)}.
\ee
In the high frequency limit, for Darwin's theory of tides, $\alpha \simeq 1$ and $\beta \simeq 1$
give $\Omega_{\rm eq}-n \propto Q^{1/2}$ and $\dot{E} \propto Q^{0}$.
In the low frequency limit, diffusion leads to $\alpha \sim 1/2$ so that
$\Omega_{\rm eq}-n \propto Q^{2/3}$ and $\dot{E} \propto Q^{1/3}$, i.e.
heating rate increases with $Q$. Lastly note that even varying $Q$ by an order of
magnitude and the heating depth by two orders of magnitude the dissipation rate varies by
less than one order of magnitude, and is in the range $\dot{E} \simeq
10^{26}-10^{28}\ {\rm erg\ s^{-1}}$ for hot Jupiters at $P_{\rm orb} \la 5\ {\rm days}$.
This heating rate is sufficient to explain the large radii of hot Jupiters, as seen by comparing
figures \ref{fig:lum_vs_rad} and \ref{fig:eqheat}.

Thermal equilibria are stable to small perturbations in $R_p$ if the perturbation to cooling dominates
that of heating. If $\dot{E}^{\rm (GT)}_{\rm heat} \propto R_p^c$ and $L_{\rm core} \propto R_p^d$,
thermal stability requires $d>c$. The tidal heating rate in eq.\ref{eq:Edotgt} has $c=4$. 
Figure \ref{fig:lum_vs_rad} shows that the cooling rate has $d \simeq 8-16$ over the 
range of radii shown, implying stability.  We find, however, that at a larger radius $R_p \ga 3R_J$, the exponent $d$ becomes less than 4, implying thermal instability.


\section{ Eccentricity evolution }
\label{sec:eccdot}

The orbits of some hot Jupiters are clearly more circular than for the
population of long period planets. This is often attributed to circularization
by the gravitational tide raised in the planet by the star. The rate of
change of eccentricity is \citep{2002ApJ...573..829M}
\be
\dot{e}^{\rm (GT)} & = & - e \left( \frac{81n}{2Q_p'} \right) 
\left( \frac{M_\star}{M_p} \right) \left( \frac{R_p}{a} \right)^5
\left( f_1(e) - \frac{11}{18} f_2(e) \frac{\Omega}{n} \right)
\label{eq:edotGT}
\ee
where 
\be
f_1(e) &= & \frac{1 + (15/4)e^2 + (15/8)e^4 + (5/64)e^6}{(1-e^2)^{13/2}}
\ee
and
\be
f_2(e) &= & \frac{1 + (3/2)e^2 + (1/8)e^4}{(1-e^2)^{5}}. 
\ee
For the equilibrium spin rate implied by $N^{\rm (GT)}=0$ (see Hut (1981), eq.11),
 the bracketed factor in eq.\ref{eq:edotGT} is positive, implying circularization. 
The circularization time is then
\be
&& t_{\rm circ, GT}   =    - \frac{e}{\dot{e}^{\rm (GT)}} = 2 \times 10^8\ {\rm yrs}\ 
\left( \frac{P_{\rm orb}}{4\ {\rm days}}\right)^{13/3}  \left( \frac{Q_p'}{10^6} \right) 
 \nonumber \\ & \times &
 \left( \frac{P_p}{4\ {\rm hrs}} \right)^{-10/3}
  \left( \frac{M_\star}{10^3M_p} \right)^{2/3}
 \left( f_1(e) - \frac{11}{18} f_2(e) \frac{\Omega}{n} \right)^{-1}.
\ee
For hot Jupiters
with radii $R=(1.0-1.8)\times R_J$, $Q_p' \leq 10^6$ and 
$P_{\rm orb} \la 5\ {\rm days}$ the orbits
should be highly circular after $5$ Gyr. While many orbits are 
consistent with zero eccentricity, it
has been emphasized by  \citet{2008ApJ...686L..29M} that nearly $1/4$ 
of planets within $0.1$AU have
$e \geq 0.1$. In the absence of additional perturbations 
to the orbit, the finite eccentricity can
only be explained by rather large $ Q_p'  \sim 10^8$. The question 
is then why planets with
presumably similar structure and orbital periods should have 
tidal Q differing by orders of magnitude.

Inclusion of both the thermal and gravitational tides changes 
this simple picture in two ways.

First, eq.\ref{eq:edotGT} shows that there is a critical spin rate
$\Omega=n(18/11)(f_1/f_2)$ above which the gravitational tide {\it
excites}, rather than damps, eccentricity. For evolution only under
the gravitational tide, such rapid spin rates occur only for young,
rapidly rotating planets. Synchronization occurs on a timescale shorter
than the circularization time by $\sim 10^3$, so that negligible
eccentricity is excited in the early spin down phase. 
However, taking into account {\it both}
gravitational and thermal tide torques, \S \ref{sec:spineq} shows
that significant deviations from synchronous rotation occur in the steady 
state, over the entire life of the system.  For $e=0$
spin equilibrium, figure \ref{fig:eqspin} shows the critical spin
frequency $\Omega=(18/11)n$. For $Q_p' =10^6$, we find the gravitational
tide can drive eccentricity for $P_{\rm orb}\ga (5-9)\ {\rm days}$,
depending on the depth of the heating layer. For $Q_p'=10^5$ the critical
orbital period is longer by a factor of a few.

Second, the thermal tide may excite eccentricity due to the fact that density {\it leads}
the forcing. The small eccentricity ($e \ll 1$), high frequency ($\sigma t_{\rm th} \gg 1$)
limit is analytically tractable. To leading order in $e$, 
the results of \S \ref{sec:highfreq} plugged into eq.\ref{eq:edot0} yield
\be
\dot{e}^{\rm (TT)} & = & e \frac{3\pi}{16} \left( \frac{R_p^4F_\star(a)}{M_pC_p T a^2} \right)
\nonumber \\ & \times & 
\left( 1 + \frac{21 n/2}{3n-2\Omega} + \frac{n/2}{2\Omega-n} + \frac{n/2}{\Omega-n} \right).
\label{eq:edotTT0}
\ee
The four terms in the last parenthesis are the 
${\mathcal O}(e)$ harmonics $(m,k)=(0,1), (2,3)$, $(2,1)$ and $(2,2)$. 
Depending on $\Omega/n$, 
eccentricity can either be excited or damped. 
Large response occur near the three resonances $P_{\rm spin}/P_{\rm orb}
=2/3,1,2$.

\begin{figure}
\plotone{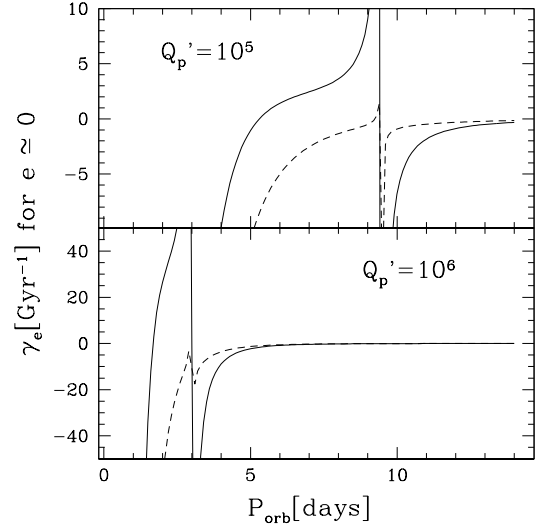}
\caption{ Growth rate of eccentricity $e$ in the $e \ll 1$ limit, using the high frequency
limit for the thermal tide (see eq.\ref{eq:gammae}) and the equilibrium spin frequencies
for $e=0$ from eq.\ref{eq:asynch}. Positive values denote eccentricity growth
and negative values denote eccentricity damping. The solid (dashed) lines are the
super-(sub-) synchronous solutions for the spin. The upper panel is for strong gravitational
tide ($Q_p'=10^5$) while the lower panel is for weak gravitational tide ($Q_p'=10^6$). 
A solar type star was used, and $M_p=M_J$ and $R_p=1.3R_J$. The divergences are due to the $(k,m)=(3,2)$ and $(1,2)$
resonances for the thermal tide terms in eq.\ref{eq:gammae}.
}
\label{fig:gammae}
\end{figure}

At small eccentricity, both the gravitational and thermal tides have $\dot{e} \propto e$, 
implying either exponential growth or decay with the rate (see eq.\ref{eq:edotGT}, \ref{eq:edotTT0}
and \ref{eq:omegaeq})
\be
\gamma_e & = & \frac{\dot{e}^{\rm (GT)} + \dot{e}^{\rm (TT)}}{e}
\nonumber \\ & = &
  \left( \frac{81n}{2Q_p'} \right) \left( \frac{M_\star}{M_p} \right) \left( \frac{R_p}{a}
\right)^5
\left[ - \left( 1 - \frac{11}{18} \frac{\Omega}{n} \right)
\right. \nonumber \\ & + & \left.
\frac{1}{9} \left( \frac{\Omega_{\rm eq}-n}{n} \right)^2
\left( 1 + \frac{21 n/2}{3n-2\Omega} + \frac{n/2}{2\Omega-n}  + \frac{n/2}{\Omega-n}  \right)
\right]
\label{eq:gammae}
\ee
Figure \ref{fig:gammae} evaluates eq.\ref{eq:gammae} using $\Omega=\Omega_{\rm eq}$ for $e=0$ from
eq.\ref{eq:asynch}, for the supersynchronous solution (solid line)
and the subsynchronous solution (dashed line). Two different tidal
$Q_p'=10^5-10^6$ are shown. The value of $Q_p'$ determines how long
the orbital period must be before the spin becomes sufficiently
asynchronous for $\gamma_e$ to change sign.  In both cases shown
in figure \ref{fig:gammae}, the same basic pattern is found. For the
supersynchronous case, close to the star eccentricity is strongly damped,
then a region of eccentricity growth ending with a resonance, then
a region of eccentricity damping starting with the same resonance. The eccentricity
is always driven at long orbital periods, although the growth rate is too small
to affect the orbit over the system's lifetime. In the regions of large
eccentricity growth, even if a planet began with small $e$,
10-100 efoldings would imply that the eccentricity could grow to large
values. For small $Q_p'=10^5$, eccentricity may be driven even out to
$P_{\rm orb}=10\ {\rm days}$, implying the observed eccentricity at these
relatively large orbital periods may not be primordial, but rather influenced by the
thermal tide. For the subsynchronous case, the regions of eccentricity
growth are far more limited, and may be confined to orbital periods very
near the resonance.

We have shown that small, but finite, eccentricity is unstable for small ``windows" in $P_{\rm orb}$.
 Hence even if all orbits began as nearly circular, a non-monotonic
distribution of eccentricity would result, in which eccentricity would be
small near the star, possibly large in the window where $\gamma_e > 0$,
and again small in the exterior region where $\gamma_e \simeq 0$.
The $P_{\rm orb}$ range for large eccentricity may differ from
one system to the next, as it depends on factors such as stellar flux and planet
mass, radius and tidal $Q_p'$, and heating depth $\kappa_\star^{-1}$.  This effect
may explain the non-monatonic distribution of observed eccentricities
for hot Jupiters.

\section{ Simultaneous spin, orbit and thermal equilibrium }
\label{sec:thermal}

\begin{figure*}
\plottwo{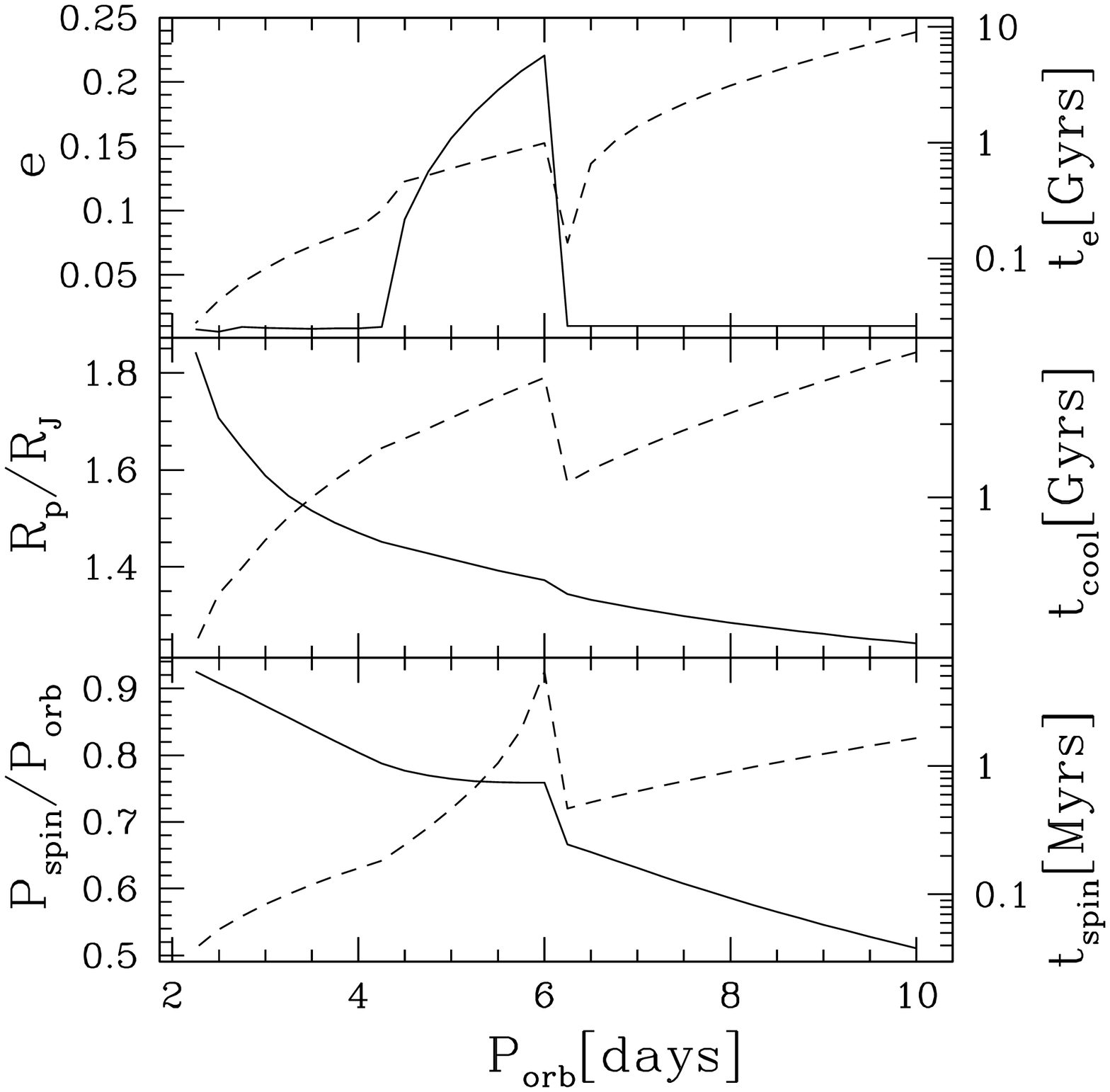}{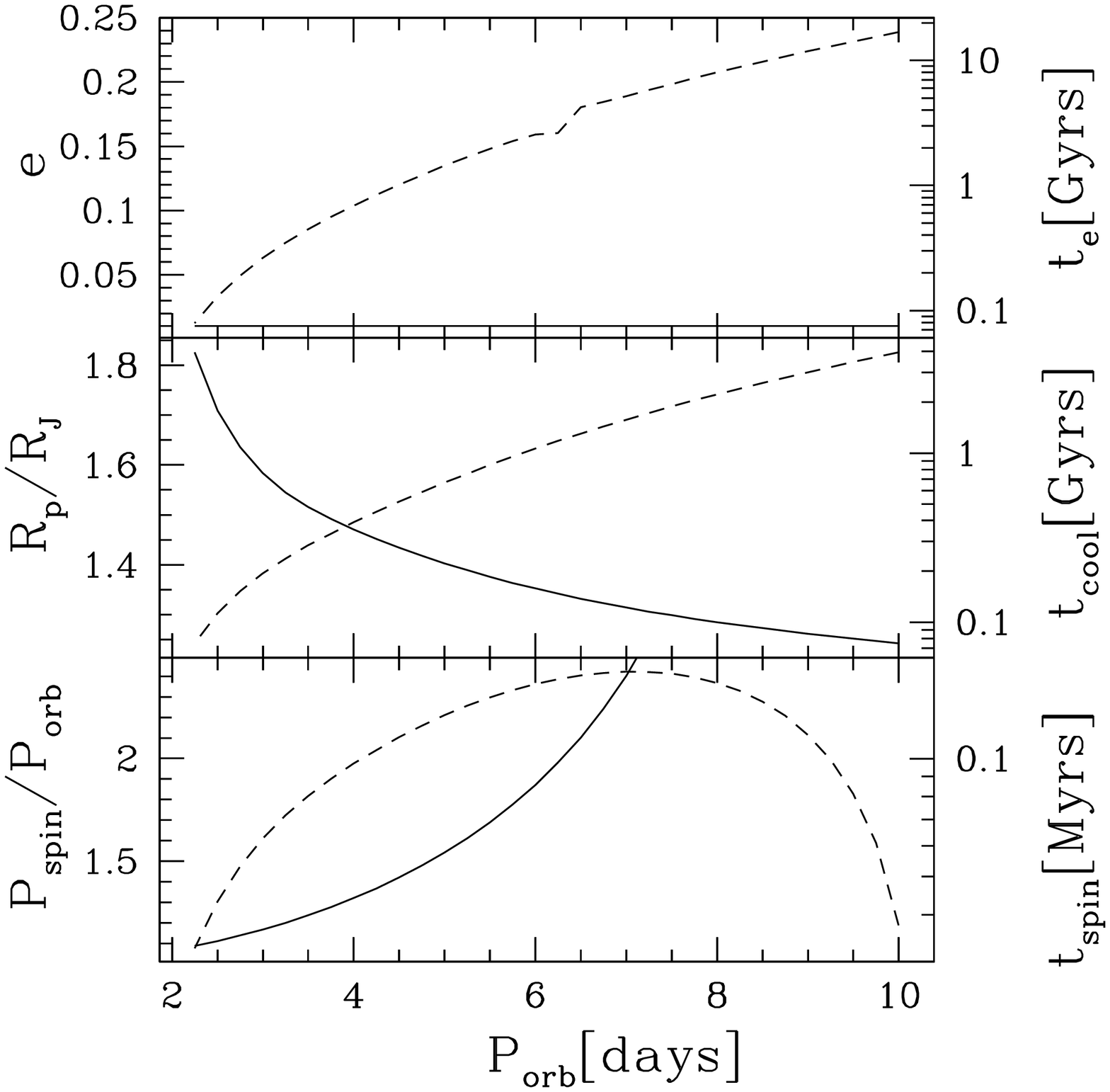}
\caption{ Equilibrium values of $P_{\rm spin}$, $R_p$ and $e$ for $M_p=M_J$, 
$\kappa_\star=10^{-2}\ {\rm cm^2\ g^{-1}}$ and $Q_p'=10^6$. The left (right) plot
shows the super- (sub-) synchronous solution. In each panel, the 
solid line is for the left axis showing $e, R_p$ and $P_{\rm spin}$, and  the dashed line
is for the right axis showing the timescales to attain equilibrium.
}
\label{fig:eqspinrade}
\end{figure*}

In this section we self-consistently solve for the values of 
$P_{\rm spin}$, $R_p$ and $e$ which satisfy 
\be
N^{\rm (TT)} + N^{\rm (GT)} & = & 0
\\
\dot{E}^{\rm (GT)}_{\rm heat} - L_{\rm core} & = & 0
\\
\dot{e}^{\rm (GT)} + \dot{e}^{\rm (TT)} & =& 0.
\ee
Rather than choosing fixed values of $F_{\rm core}$ independent of latitude, we
follow appendix \ref{sec:appendix2} to compute $F_{\rm core}(\theta)$ by requiring
that the radiative atmosphere match smoothly to an adiabatic interior with
the correct $M_p$ and $R_p$. Given this background solution for the atmosphere,
we solve the boundary value problem for the linearized perturbations (see \S
\ref{sec:linear}). This includes diffusion explicitly, and interpolates between
the high and low frequency limits. The perturbations are then used to 
compute the quadrupole moments by integration over column (eq.\ref{eq:sigmalin})
 and latitude (eq.\ref{eq:Qlm2}). Enough terms are used in the frequency sums
in eq.\ref{eq:edotsec}, \ref{eq:torquesec} and \ref{eq:heatrate} to ensure convergence.
Newton's method is used find equilibrium values given a guess at the solution.
We choose a small starting value for eccentricity, $e=0.01$. If $\gamma_e>0$, the eccentricity
grows until it reaches a finite equilibrium value. If $\gamma_e<0$ at $e=0.01$, we treat $e=0$ as a
good equilibrium solution.

Results are shown in figure \ref{fig:eqspinrade} for $M_p=M_J$, 
$\kappa_\star=10^{-2}\ {\rm cm^2\ g^{-1}}$ and $Q_p'=10^6$. 
First we discuss the super-synchronous solution in the left panel.
Close to the star,
the gravitational tide dominates, resulting in nearly synchronous spin and circular
orbit. The tidal heating rate due to asynchronous spin grows rapidly toward the star, leading to an increase in radius.
The numerical values can be understood by comparing figures \ref{fig:lum_vs_rad} and \ref{fig:eqheat}.
In between $P_{\rm orb} \simeq 4-6\ {\rm days}$, $\gamma_e>0$ and $e=0$ is not a stable equilibrium.
We find finite equilibrium eccentricity with values $e=0.0-0.2$. Finite eccentricity does not
seem to be as important as asynchronous spin in the tidal heating rate since no feature
appears between 4-6 days for $R_p$. The eccentricity equilibria in figure \ref{fig:eqspinrade} are stable. For the $e=0$ equilibria, $de/dt<0$ implies stability. In the window of instability, $de/dt>0$ for 
$e<e_{\rm eq}$ while $de/dt<0$ for $e>e_{\rm eq}$.

The timescale to return to spin equilibrium for a small deviation from equilibrium is
\be
t_{\rm spin} & \simeq \frac{I_p\Omega_{\rm eq}}{N_{\rm eq}}
\ee
where $N_{\rm eq}=|N^{\rm (TT)}_{\rm eq}|=|N^{\rm (GT)}_{\rm eq}|$ is the value of the opposing
torques in equilibrium. This timescale is shorter than the age by a factor $\sim 10^3$ for
$P_{\rm orb} \la 1$ week. The core cooling time in eq.\ref{eq:tcool} is comparable to or shorter
than the Gyr age for $P_{\rm orb} \la 10$ days. The timescale to reach eccentricity
equilibrium is
\be
t_{\rm e} & = & \left( \frac{d\ln e_{\rm eq}}{dt} \right)^{-1}
\ee
where $d\ln e_{\rm eq}/dt=|d\ln e^{\rm (GT)}_{\rm eq}/dt|=|d\ln e^{\rm (TT)}_{\rm eq}/dt|$,
evaluated at equilibrium. The eccentricity evolution time is also comparable to or shorter than
the age inside 1 week. Close to the star, we expect all three quantities to have equilibrium values,
while further away there may not be sufficient time for $R_p$ and $e$ to attain equilibrium values.

The sub-synchronous solution is shown in the right hand panel of figure \ref{fig:eqspinrade}. While the radius is similar to the that of the super-synchronous solution, the spin and eccentricity differ. The $(m,k)=(2,1)$ resonance is weaker than the $(m,k)=(2,3)$ resonance due to the prefactor in eq.\ref{eq:edotTT0}, and eccentricity is driven less strongly
than the super-synchronous case. At the evenly spaced orbital periods used, 
we did not get close enough to the resonance for the driving to be significant. 
The spin period becomes large for long orbital period, and will eventually turn into retrograde spin.

Although not plotted in figure \ref{fig:eqspinrade}, we found additional finite eccentricity solutions
for $P_{\rm orb} \ga 6$ days, where $e=0$ is also a solution. We have not thoroughly investigated
large $e$ equilibria which cannot be found from a small starting guess for $e$. We leave this for a
future investigation, but note that many equilibria may be possible due to higher order thermal
tide resonances that become important for large $e$. The ability to trap the orbit into one of these
resonances during evolution will require a more detailed numerical investigation.

We now discuss how the results in figure \ref{fig:eqspinrade} would change for different parameters.
From figure \ref{fig:lum_vs_rad}, increasing $M_p$ leads to larger
cooling rates at fixed $R_p$. To bring heating ($\propto R_p^4$) and cooling ($\propto M_p R_p^{8-16}$) back into balance
requires that higher mass planets must have smaller radii, and vice versa for lower mass planets. For smaller $Q_p'$
(larger gravitational tide dissipation rate), the planet must be further from the star before significant 
asynchronous spin develops. This pushes the strong resonance driving eccentricity further
from the star (see figure \ref{fig:gammae}), with a drop in the magnitude of the driving rate.
Increasing $F_\star(a)$ and $e$, and decreasing $\kappa_\star$, acts to increase the strength of the 
thermal tide, pushing the spin away from synchronous and the region of $\gamma_e>0$ closer to the star where $\gamma_e$ is larger.

Our calculations are successful from the point of view that we have identified a powerful internal heat source
which may explain the large radii observed for the transiting planets. However, our theory 
predicts a correlation of radius with orbital period, and nonzero eccentricity over a small
range in orbital period.
The plots of $R_p$ and $e$ versus $P_{\rm orb}$ for the observed transiting planets do not clearly show the
simple behavior found in figure \ref{fig:eqspinrade}. 
We speculate that part of the variance is due to differences in planet mass, stellar flux and heating depth
($\kappa_\star$) associated with the TiO/VO transition. Furthermore, our radii are upper limits, as we do
not include a core. We intend to make more detailed comparison to data in a future investigation.

\section{ Differential rotation}
\label{sec:diffrot}

In this section we discuss the wind profiles induced by the thermal tide
torque and estimate dissipation rates due to differential rotation.

The time-averaged force per unit volume on a fluid element due to the stellar
tidal force acting on the thermal tide is
\be
\vec{f}(\vec{x}) & = & - \langle \delta \rho \grad U \rangle_t
= \rho \langle \frac{\delta T}{T} \grad U \rangle_t.
\ee
In the high frequency limit, for heating function $\delta \epsilon
= \kappa_\star F_\star \exp(-\kappa_\star y)$  we find a zonal component
\be
f_\phi & \simeq &  \rho n^2 R_p \frac{F_\star \kappa_\star}{(\Omega-n) C_p T}
e^{-\kappa_\star y}.
\label{eq:fphi}
\ee
The ratio of the zonal acceleration $a_\phi=f_\phi/\rho$ on a fluid element to the planet's gravity is then
\be
\frac{a_\phi}{g} & \sim & \frac{n^2 R_p}{g} (\sigma t_{\rm th})^{-1}
\sim \frac{M_\star}{M_p} \left( \frac{R_p}{a} \right)^3 (\sigma t_{\rm th})^{-1}
\nonumber \\ & \sim & 
10^{-3} \left( \frac{M_\star}{10^3 M_p} \right) \left( \frac{100\ R_p}{a} \right)^3
(\sigma t_{\rm th})^{-1}.
\ee
In the absence of friction and Coriolis forces, the torqued layers would
accelerate to the sound speed on a timescale of days. Coriolis forces
will act to bend the zonal winds to produce circulation patterns with
characteristic velocities $v_{\rm cir} \sim a_\phi/\Omega \cos\theta$
sufficiently far from the equator.  More importantly, the thermal
tide force deposits net angular momentum. In the likely event that
opposing gravitational tidal torques are exerted in a deeper layer,
angular momentum transport is required to connect the source, due to
the thermal tide, and the sink due to the gravitational tide.

As a simplest possible model, we ignore the Coriolis force and consider a turbulent
boundary layer model (e.g. \citealt{ll87}) that is forced by 
the thermal and gravitational tidal torques. The thermal tide exerts a stress
\be
\tau & = & \int dz f_\phi
= \int dy\ n^2 R_p \frac{F_\star \kappa_\star}{(\Omega-n) C_p T}
e^{-\kappa_\star y}
\nonumber \\ & = &
n^2 R_p \frac{F_\star}{(\Omega-n) C_p T}
\label{eq:tau}
\ee
on the tidally forced layer. We treat this stress as applied in a thin layer at
$z=z_1$, and the gravitational tide exerts an opposite stress
at $z=z_2<z_1$, deeper in the layer. The horizontal momentum equation
for our toy model with horizontal velocity $v$ and kinematic viscosity $\nu$ is
\be
0 & = & \rho \nu \frac{d^2v}{dz^2} + \tau \delta (z-z_1) - \tau \delta (z-z_2).
\ee
Setting the shear to be zero outsize $z_2 \leq z \leq z_1$, we find the shear
in between the two forced layers to be
\be
\frac{dv}{dz} & = & \frac{\tau}{\rho \nu}.
\label{eq:shear}
\ee
Turbulent velocity fluctuations of size
\be
v_\star &\sim &(\tau/\rho)^{1/2}
= 40\ {\rm m\ s^{-1}}\  \left( \frac{10^{-5}\ {\rm g\ cm^{-3}}}{\rho} \right)^{1/2}
\nonumber \\ & \times & 
\left( \frac{n}{\Omega-n} \right)^{1/2}
\left( \frac{R_p}{10^{10}\ {\rm cm}} \right)^{1/2}
\left( \frac{4\ {\rm days}}{P_{\rm orb}} \right)^{3/2}
\ee
transport momentum. We take a characteristic vertical
distance of a scale height $H$, so that the viscosity $\nu \sim H v_\star$.
Eq.\ref{eq:shear} then becomes $dv/dz \sim v_\star/H$. The shearing implies an energy 
dissipation rate
\be
\dot{E}_{\rm shear} & \sim & 4\pi R_p^2 \int dz \rho \nu \left( \frac{dv}{dz} \right)^2
\sim 4\pi R_p^2 \rho v_\star^3
\nonumber \\ & = & 
8 \times 10^{27}\ {\rm erg\ s^{-1}}
\left( \frac{10^{-5}\ {\rm g\ cm^{-3}}}{\rho} \right)^{1/2}
\nonumber \\ & \times &
\left( \frac{n}{\Omega-n} \right)^{3/2}
\left( \frac{R_p}{10^{10}\ {\rm cm}} \right)^{7/2}
\left( \frac{4\ {\rm days}}{P_{\rm orb}} \right)^{9/2}.
\ee
This estimate for dissipation due to shearing is comparable 
to the dissipation rate of the gravitational tide in eq.\ref{eq:Edotgt}.

\section{ Summary and conclusions }
\label{sec:summary}

We investigated the role of thermal tides in short-orbital
period gas giant exoplanets, the hot Jupiters. Our investigation consisted
of two parts. We computed the time-dependent temperature and
density perturbations in radiative atmospheres. We then applied
these results for the thermal tide, and the Darwin-Hut theory of the
gravitational tide, to compute spin, radius and eccentricity evolution.

Previous investigations (e.g. \citealt{1978Natur.275...37I}) for the
case of Venus considered a shallow atmosphere bounded by ground at the
base. In considering the thermal tide in hot Jupiters, we used
radiative heat transport, and allowed incoming stellar radiation to be
fully absorbed in the atmosphere.  We computed the temperature
perturbations in the atmosphere resulting
from time-dependent insolation in order to estimate the induced
thermal tide quadrupole moment.  As shown in in the context of
Venus, in the high frequency limit, the thermal tidal bulge {\it
leads} the subsolar point \citep{1978Natur.275...37I}. In considering
deep atmospheres, we include the effects of thermal diffusion, and
found that the thermal tidal bulge still leads the substellar point
for low tidal forcing frequencies, when heat can diffuse
deep into the atmosphere.  This result for the phase lead (for
density, lag for temperature) is simplest to see in the results of
nonlinear simulations in figures
\ref{fig:TP_fullday_deep_Porb=4_Pspin=3}
 - \ref{fig:colvst}. Downward going diffusion waves at the base of the
heated layer are discernible in figures
\ref{fig:TP_fullday_deep_Porb=4_Pspin=3} and
\ref{fig:linear_shallow_lowfreq} for the low frequency limit. The
effective column and quadrupole moment are summarized in analytic
formula in the high and low frequency limits in
eq.\ref{eq:Sigmamkhighfreq}, \ref{eq:TTQlm_highfreq},
\ref{eq:sigmalowfreq} and \ref{eq:fit}.

In appendix \ref{sec:appendix1}, we derive the evolutionary equations for the planet's
spin and orbit, as well as tidal heating rates for arbitrary
quadrupole moments. These formula as used with the thermal tide
quadrupole to compute equilibrium spin rates (\S
\ref{sec:spineq}), equilibrium tidal heating rates (\S
\ref{sec:radii}), the growth/damping rate of eccentricity
(\S \ref{sec:eccdot}), and simultaneous spin, orbital and
thermal equilibrium (\S \ref{sec:thermal}).

We find, contrary to the widely held assumption, that for reasonable
heating depths and tidal $Q$, {\it hot Jupiters are far from
a synchronous spin state} (see figure \ref{fig:eqspin}). At long
orbital periods, the thermal tide torque dominates
and the planet is highly asynchronous, while at smaller radii
the gravitational tidal torque is larger, and the planet approaches
a synchronous state. Direct measurements of the planetary rotation
rate could confirm asynchronous rotation, although this appears
difficult in the near future. Possible methods include
detection of centrifugal flattening using high photometric precision
transit observations \citep{2002ApJ...574.1004S}, and Doppler shifted
absorption lines for transiting planets \citep{2007ApJ...669.1324S}.

A more promising route to observationally constrain rotation rates at
the moment may be through the indirect effect of planetary rotation on
zonal winds, and the resulting change in phase of maximum thermal emission
in Spitzer infrared observations.  We have found, for uniform rotation,
that flux perturbations are smaller in magnitude and have longer lag times
in the high frequency limit than in the low frequency limit (see figure
\ref{fig:Fvst}). The strength of zonal winds is expected to decrease
for more rapid rotation rate as well \citep{2008arXiv0809.2089S}.
Hydrodynamic simulations of global circulation already contain the thermal
tide quadrupole moment. ``Turning on" the stellar tidal force in these
simulations will torque the surface layer as described in \S
\ref{sec:diffrot}, leading to spinup of the atmosphere, as well as
momentum-driven circulation pattens in addition to the usual thermally-driven
circulation. As argued by \citet{2008arXiv0810.1282G}, a source of
dissipation such as that due to the gravitational tide or viscous friction
is necessary to damp kinetic energy added to the atmosphere, and may affect
steady-state wind speeds.

In \S \ref{sec:radii}, gravitational tidal dissipation due to
asynchronous spin is shown to yield large {\it steady state} heating
rates, sufficient in magnitude to power the large observed radii over the
range $R_p=(1.2-1.8)\ \times  R_J$ if the heat is deposited in the convective core.
\S \ref{sec:diffrot} shows that vertical shearing in the wind profiles
between the layers torqued by the thermal and gravitational tides may also
give heating rates large enough to alter the atmospheric temperature profiles,
and perhaps slow the cooling of the planet \citep{2001ApJ...548..466B}.

In \S \ref{sec:eccdot} we show that the thermal tide is capable of
driving eccentricity. We compute the growth rate of eccentricity, in
the limit of small eccentricity, finding narrow windows in orbital period
in which the eccentricity is driven, while outside these windows it is
even more strongly damped than for the gravitational tide alone. There are
two consequences. As pointed out by \citet{2008arXiv0808.3724M}, a puzzle
exists as to why {\it all} hot Jupiter orbits are not highly circular. It
is observed that $\sim 25\%$ of hot Jupiters have eccentricities $e \geq
0.1$. \citet{2008arXiv0808.3724M} argue that this must imply the planets
with eccentric orbits have far larger $Q$ than planets with circular
orbits. Including the thermal tide, we find that it is expected that
narrow windows in orbital period should exist in which eccentricity
is driven to large values. This may explain the puzzling mix of zero
and nonzero eccentricities in the observed planets. As a corollary,
one cannot simply constrain the gravitational tidal $Q$, since it's
effects are mixed in with the thermal tide effect. In some ranges of
orbital period they combine to cause more rapid circularization, while in
other orbital period ranges they partially cancel. A more detailed study
including both effects is required to place meaningful constraints on $Q$.

\S \ref{sec:thermal} shows that the timescales to achieve spin, orbital
and thermal equilibrium are small or comparable to the Gyr ages of observed transiting
planets. Simultaneous equilibria for spin period, radius and eccentricity are found.
Radii increase strongly toward the star, mainly due to asynchronous spin, inside orbital
periods of 1-2 weeks. Asynchronous spin again increases to longer orbital periods. 
Nonzero equilibrium values of eccentricity are found in the region where $e=0$ is unstable.

In summary, this initial investigation has found that thermal tides have a significant impact
on hot Jupiter rotation rates, eccentricities and thermal state, and that 
seemingly unrelated observation puzzles, such as large radii and nonzero eccentricities, may have
a natural explanation within this model.


\acknowledgements

We thank Lars Bildsten, Peter Goldreich and Jonathan Mitchell for useful discussions. P.A. received
support from an Alfred P. Sloan fellowship, and the Fund for Excellence
in Science and Technology fellowship from the University of Virginia.  A.S. 
acknowledges support from a Lyman Spitzer Jr. Fellowship given by
Astrophysical Sciences at Princeton University as well as a Friends of 
the Institute Fellowship at the Institute for Advanced Study, in Princeton, NJ.


\appendix

\section{Spin, orbit and thermal evolution rates for eccentric orbits}
\label{sec:appendix1}

The interaction Hamiltonian coupling the density field of the planet to 
the gravitational tidal potential of the star is given by 
\be
H & = & \int d^3x \rho(\vec{x},t) U(\vec{x},t)
\label{eq:H}
\ee
where the integral is over the perturbed body of the planet, $\rho(\vec{x},t)$ 
is the time-dependent density in the planet and $ U(\vec{x},t) $ the tidal 
potential. To evaluate this expression, we expand the tidal potential in spherical 
harmonics
\be
U(\vec{x},t) & = & - GM_\star \sum_{\ell m} \frac{4\pi}{2\ell +1} 
\left( \frac{r^\ell}{D^{\ell+1}} \right)
Y_{\ell m}^*(\pi/2,\Phi) Y_{\ell m}(\theta,\phi)
\label{eq:U}
\ee
and define the time-dependent multipole moments of the planet
\be
{\cal Q}_{\ell m}(t) & =& \int d^3x\ r^{\ell} Y^*_{\ell m}(\theta,\phi) \rho(\vec{x},t).
\ee
Since $r^\ell \rho(\vec{x},t)$ is a real quantity, 
and $Y_{\ell m}^*=(-1)^m Y_{\ell, -m}$, the moments must satisfy
${\cal Q}_{\ell m}^* = {\cal Q}_{\ell, -m} (-1)^m$.
By defining $W_{\ell m}\equiv[4\pi/(2\ell+1)]Y_{\ell m}(\pi/2,0)$, the interaction 
Hamiltonian in 
eq.\ref{eq:H} may be conveniently expressed in terms of a sum over the tidal 
(spherical) harmonics
\be
H(D,\Phi) & = & -GM_\star \sum_{\ell m} W_{\ell m} 
{\cal Q}^*_{\ell m}(t) \frac{e^{-im\Phi}}{D^{\ell+1}}.
\ee
The radial and tangential accelerations on the relative motion about the center of mass are then
\be
a_D & = & - \frac{1}{\mu} \frac{\partial H}{\partial D}
\\
a_\Phi & = & - \frac{1}{\mu D} \frac{\partial H}{\partial \Phi},
\ee
where $\mu=M_pM_\star/(M_p+M_\star) \simeq M_p$ is the reduced mass.
The change in orbital energy and angular momentum are then
\citep{1999ssd..book.....M}
\be
\dot{E}_{\rm orb} & = &  \mu \left( \dot{D} a_D + D \dot{\Phi} a_\Phi \right)
\label{eq;Eorbdot}
\\
\dot{L}_{\rm orb} & = & \mu D a_\Phi.
\label{eq:Lorbdot}
\ee
Since $E_{\rm orb}=-GM_pM_\star/2a$ 
and $L_{\rm orb}=\mu \sqrt{ G(M_p+M_\star)a(1-e^2) }$,
the changes in semi-major axis and eccentricity are
\be
\frac{\dot{a}}{a} & = & \frac{2}{n a \sqrt{1-e^2}} \left[ e \sin \Phi a_D
+ (1+e\cos\Phi) a_\Phi \right]
\label{eq:adot}
\\
\frac{e\dot{e}}{1-e^2} & = & \frac{\dot{a}}{2a} - \frac{D a_\Phi}{n a^2 \sqrt{1-e^2}}.
\label{eq:edot}
\ee
The torque on the planet is
\be
N & = & - \int d^3x \rho(\vec{x,t}) \frac{\partial U(\vec{x},t)}{\partial \phi}
= \int d^3x \rho(\vec{x,t}) \frac{\partial U(\vec{x},t)}{\partial \Phi}
= - \dot{L}_{\rm orb} 
\label{eq:NGT}
\ee
since the longitudes appear exclusively in the combination $\phi-\Phi$. 
This explicitly shows angular momentum is conserved over the orbit plus planet.

Given the quadrupole moments ${\cal Q}_{\ell m}(t)$, eq.\ref{eq:adot},
\ref{eq:edot} and \ref{eq:NGT} can be integrated to find the changes
in $a$, $e$ and $\Omega$. To isolate the secular evolution, we treat
the orbit as Keplerian with elements $a$ and $e$ that vary on
timescales much longer than the orbital period. We then expand all
quantities in a Fourier series in time. Let
\be
\left( \frac{a}{D} \right)^{\ell+1} e^{-im\Phi} & = & \sum_{k=-\infty}^\infty X^{\ell m}_k(e) e^{-iknt}
\label{eq:ftorb}
\ee
where the Hansen coefficients $X^{\ell m}_k(e)$ are defined as \citep{1999ssd..book.....M}
\be
X^{\ell m}_k(e) & = & \frac{n}{2\pi} \int_0^{2\pi/n} dt e^{iknt-im\Phi} \left( \frac{a}{D} \right)^{\ell+1}
\nonumber \\ & \simeq & 
\delta_{mk} + \frac{e}{2} \left[ (l+1+2m)\delta_{k,m+1} + (l+1-2m)\delta_{k,m-1} \right] 
+ {\mathcal O}(e^2).
\label{eq:Hansen}
\ee
The Hansen coefficients are real, and satisfy $X^{\ell, -m}_{-k}=X^{\ell m}_k$. The multipole moments are similarly expanded as
\be
{\cal Q}_{\ell m}(t) & = & \sum_{k=-\infty}^\infty {\cal Q}_{\ell m k} e^{-ikn t}
\label{eq:ftQlm}
\ee
where
\be
{\cal Q}_{\ell m k} & = & \frac{n}{2\pi} \int_0^{2\pi/n} dt\ e^{iknt} Q_{\ell m}(t),
\ee
which satisfy ${\cal Q}_{\ell ,-m,-k}=(-1)^m {\cal Q}_{\ell m k}^*$.

Plugging eq.\ref{eq:ftorb} and \ref{eq:ftQlm} into eq. \ref{eq:adot}, \ref{eq:edot} and \ref{eq:Lorbdot}, using the reality conditions relating $(m,k)$ to $(-m,-k)$, and and isolating the secular terms, we find
\be
\frac{\dot{a}}{a} & = & \frac{\dot{E}_{\rm orb}}{|E_{\rm orb}|}
= n \sum_{\ell m k} W_{\ell m} X^{\ell m}_k  
\left( \frac{{\rm Im} ({\cal Q}_{\ell m k})}{M_p a^\ell} \right) \left( -2k \right)
\label{eq:adotsec}
\\
\frac{e \dot{e}}{\sqrt{1-e^2}} & =& n \sum_{\ell m k} W_{\ell m} X^{\ell m}_k 
\left( \frac{{\rm Im} ({\cal Q}_{\ell m k})}{M_p a^\ell} \right) \left( m- k\sqrt{1-e^2} \right)
\label{eq:edotsec}
\\
N& = & - \dot{L}_{\rm orb}  
= \mu n^2 a^2  \sum_{\ell m k} W_{\ell m} X^{\ell m}_k 
\left( \frac{{\rm Im} ({\cal Q}_{\ell m k})}{M_p a^\ell} \right) \left(  m \right).
\label{eq:torquesec}
\ee
The change in spin energy of the planet is
\be
\dot{E}_{\rm spin} & = & \Omega N 
= \mu n^2 a^2 \Omega  \sum_{\ell m k} W_{\ell m} X^{\ell m}_k 
\left( \frac{{\rm Im} ({\cal Q}_{\ell m k})}{M_p a^\ell} \right) \left(  m \right).
\ee
The energy change in the orbit and spin is given by
\be
\dot{E}_{\rm orb} + \dot{E}_{\rm spin} & = & -
\mu n^2 a^2 \sum_{\ell m k} W_{\ell m} X^{\ell m}_k  
\left( \frac{{\rm Im} ({\cal Q}_{\ell m k})}{M_p a^\ell} \right) \left(  kn-m\Omega \right).
\label{eq:Edotorbspin}
\ee
Here $\sigma_{km}=kn-m\Omega$ is the forcing frequency in the frame corotating with the planet.

In the case of gravitational tides whose phase lag is due to dissipation, energy is taken out of 
the spin or orbit and deposited as heat in the planet, so $\dot{E}_{\rm orb} + \dot{E}_{\rm
spin}<0$. In this case we define the heating rate of the planet as
\be
\dot{E}^{\rm (GT)}_{\rm heat} & = & - \left( \dot{E}_{\rm orb} + \dot{E}_{\rm
spin} \right)^{\rm GT}
= \mu n^2 a^2 \sum_{\ell m k} W_{\ell m} X^{\ell m}_k
\left( \frac{{\rm Im} ({\cal Q}_{\ell m k}^{\rm (GT)})}{M_p a^\ell} \right) \left(  kn-m\Omega \right)
\label{eq:heatrate}
\ee
This heating rate will be used for the thermal evolution of the
planet.  In the case of the thermal tide, work was needed to move the 
fluid against the stellar tidal force, so that $\dot{E}_{\rm orb} +
\dot{E}_{\rm spin}>0$. Ultimately this energy is derived from the
stellar radiation field. In this case we define the rate of
work done on the atmosphere due to heating from insolation as
\be
\dot{E}^{\rm (TT)}_{\rm work} & = & \left( \dot{E}_{\rm orb} + \dot{E}_{\rm
spin} \right)^{\rm TT}
= -\mu n^2 a^2 \sum_{\ell m k} W_{\ell m} X^{\ell m}_k
\left( \frac{{\rm Im} ({\cal Q}_{\ell m k}^{\rm (TT)})}{M_p a^\ell} \right) 
\left(  kn-m\Omega \right).
\label{eq:workrate}
\ee 
Since we do not include zonal motions in the equations
solved in this paper, we cannot explicitly compute this work done
against gravity. Our assumption is that the constant pressure
approximation adequately represents the physics of heating by
insolation. We will include fluid motions and compute this work in a
future investigation.

In equilibrium, the work done on the atmosphere first goes into spin
or orbital energy, and then is dissipated as heat in the
planet. Before equilibrium is reached, energy can be either lost or
gained from the spin and orbit.

The secular equations depend on the component of the quadrupole moments out of
phase with the tidal acceleration, hence the imaginary components are required in the
eq.\ref{eq:adotsec}, \ref{eq:edotsec} and \ref{eq:torquesec}. While we perform the sums
over $\ell$, $m$ and $k$ until convergence is achieved for our numerical work, it is useful
to have analytic limits to compare to.
The torque on the planet for a circular orbit is dominated by the
semi-diurnal term with $|m|=2$ and forcing frequency $2(n-\Omega)$. The torque
at this order is
\be
N & = & 4 \left( \frac{3\pi}{10} \right)^{1/2} \left( \frac{M_p+M_\star}{M_\star} \right) n^2
{\rm Im} \left( {\mathcal Q}_{222} \right).
\label{eq:Ntt2}
\ee
The sign in eq.\ref{eq:Ntt2} is such that density perturbations {\it leading}
the heating tend to torque the planet {\it away} from synchronous rotation,
and vice versa.
A schematic drawing for  the relevant semi-diurnal tide ($m=2$) is shown in
fig. \ref{fig:torque_schematic}.

At lowest order in $e$, the circularization rate is dominated by the $m=0$ harmonic
with frequency $n$, as well as the $|m|=2$ harmonics with frequencies
$2(n-\Omega)$, $3n-2\Omega$, and $n-2\Omega$. The circularization rate to 
lowest order in $e$ is
\be
\dot{e} & = & \sqrt{ \frac{\pi}{5} }\ \frac{n}{M_p a^2}
\left[
\sqrt{6}e\ {\rm Im} \left( {\mathcal Q}_{222} \right)
+ 3\ {\rm Im} \left( {\mathcal Q}_{201} \right)
- 7 \sqrt{ \frac{3}{2} }\  {\rm Im} \left( {\mathcal Q}_{223} \right)
- \sqrt{ \frac{3}{2} }\  {\rm Im} \left( {\mathcal Q}_{221} \right)
\right].
\label{eq:edot0}
\ee
Eccentricity may be either driven or damped depending on the relative sizes and signs of each term.

\section{ Mass-radius relation and core cooling luminosity }
\label{sec:appendix2}

Our model consists of an adiabatic interior connected to the ``background" atmosphere structure described in                              
\S \ref{sec:linear}. We do not include a solid core at the center, which would decrease the radius.

For the adiabatic interior we solve
\be
\frac{dm}{dr} & = & 4\pi r^2 \rho 
\\
\frac{dP}{dr} & = & - \frac{Gm\rho}{r^2}
\\
S(P,T) & = & S_{\rm core}={\rm constant}.
\ee
We use the equation of state from \citet{1995ApJS...99..713S} with 70\% hydrogen and 30\% helium by mass.
Given the core entropy $S_{\rm core}$ and central pressure $P_c$, the equations are integrated out
to a reference pressure $P_{\rm ref}=10^4\ {\rm dyne\ cm^{-2}}$, yielding a mass $M_p$,
radius $R_p$, and surface gravity $g=GM_p/R_p^2$. Note that the computed radii are at the fiducial
pressure, not the self-consistent radius at the optical or infrared photosphere including transit geometry
effects. The error in the radius will be of order a few scale heights, which amounts to of order a percent
of the planetary radius.

Given the solution for the adiabatic core, we integrate eq.\ref{eq:bgdTdy} and \ref{eq:bgF}
inward from the top of the atmosphere. In addition to ``given" parameters $F_*(a)$, $e$, $\theta$ and $\kappa_*$, a guess for 
$F_{\rm core}$ must be chosen. We integrate inward until the Schwarszchild criterion is violated, yielding
the entropy of the radiative atmosphere, $S_{\rm atm}$, at the radiative-convective boundary. In order that 
the radiative atmosphere matches smoothly onto the adiabatic interior, we adjust $F_{\rm core}$ until
$S_{\rm atm}=S_{\rm core}$. This yields the ``background" radiative atmosphere, $T$ versus $P$, used to find the 
perturbations in \S \ref{sec:linear}. The cooling luminosity of the core is found by integrating
$F_{\rm core}$ over latitude
\be
L_{\rm core} & = & 2\pi R_p^2 \int_0^\pi d\theta \sin\theta\  F_{\rm core}(\theta).
\label{eq:Lcore}
\ee
An example of $L_{\rm core}$ is given in figure \ref{fig:lum_vs_rad}.


This model for the mass-radius relation and cooling luminosity is a modified version \citet{2006ApJ...650..394A}.
Their upper boundary condition assumed the existence of an isothermal region just below where the stellar radiation
is absorbed. They assigned this region the temperature $T_{\rm deep}$, and treated this as a parameter of the model.
Here we improve on \citet{2006ApJ...650..394A} by calculating, rather than parametrizing, the temperature at the atmosphere.
 

\end{document}